\documentclass{book}
\usepackage{psfrag}
\usepackage{makeidx}
\usepackage{amsthm,amsmath,amssymb,amsfonts}
\usepackage{setspace,graphicx}
\usepackage{Generic}  
\usepackage[sort,longnamesfirst]{natbib}
\usepackage{url} 
\usepackage{todonotes}

\usepackage{caption}
\usepackage{subcaption}
\usepackage{rotating}
\usepackage[normalem]{ulem}

\newcommand{\pcite}[1]{\citeauthor{#1}'s \citeyearpar{#1}}
\numberwithin{equation}{section}
\theoremstyle{plain}
\newtheorem{theorem}{Theorem}[section]

\begin{document}
\newcommand{\bm}[1]{\boldsymbol{#1}}
\newcommand{\mc}[1]{\mathcal{#1}}
\newcommand{\Data}{\mc{D}}
\newcommand{\bth}{\bm{\theta}}

\newcommand{\laurence}[1]{\textcolor{purple}{{\bf LD}:\ #1}}
\newcommand{\yanan}[1]{\textcolor{red}{\ #1}}
\newcommand{\scott}[1]{\textcolor{blue}{{\bf SS}:\ #1}}
\doublespacing
%


\chapter[Reversible Jump MCMC]{Reversible jump Markov chain Monte Carlo and multi-model samplers}


\begin{center}
\begin{large}
{\em Yanan Fan, Scott A. Sisson  and Laurence Davies}
\end{large}
\end{center}


\section{Introduction} 
\label{s:intro}


The  reversible jump Markov chain Monte Carlo (RJMCMC) sampler
\citep{green95} provides a general framework for Markov chain Monte Carlo (MCMC) simulation in which the dimension of the parameter space can vary between iterates of the Markov chain. The reversible jump sampler can be viewed as an extension of the Metropolis-Hastings algorithm onto more general state spaces.

To understand this in a Bayesian modelling context, suppose that for observed data $\Data$ we have a countable collection of candidate models ${\cal M} = \{ {\cal M}_1, {\cal M}_2, \ldots \}$ indexed by a parameter $k  \in {\cal K}$.  The index $k$ can be considered as an auxiliary model indicator variable, such that ${\cal M}_{k'}$ denotes the model
where $k=k'$.
Each model ${\cal M}_k$ has an $n_k$-dimensional 
vector of unknown parameters, $\boldsymbol{\theta}_k \in {\cal R}^{n_k}$, where 
$n_k$ can take different values for different models $k\in\mathcal{K}$. The joint  posterior distribution of 
$(k, \boldsymbol{\theta}_k)$ given observed data, $\Data$, is obtained as the product of the likelihood, 
$L(\Data| k,\boldsymbol{\theta}_k)$, and the joint prior, $p(k,\boldsymbol{\theta}_k)=p(\boldsymbol{\theta_k}| k)p(k)$, constructed from the prior distribution
of $\boldsymbol{\theta}_k$ under model ${\cal M}_k$, and the prior for the model indicator $k$ (i.e.~the prior for model $\mathcal{M}_k$). Hence the joint posterior is
\begin{equation}\label{eqn:jointpost}
	\pi(k, \boldsymbol{\theta}_k | \Data) = \frac{
	L(\Data | k, \boldsymbol{\theta}_k)p(\boldsymbol{\theta}_k| k)p(k)}{
	\sum_{k' \in {\cal K}}  \int_{{\cal R}^{n_{k'}}} L(\Data | k', \boldsymbol{\theta}'_{k'})p(\boldsymbol{\theta}'_{k'}| k')p(k') d\boldsymbol{\theta}_{k'}'}.
\end{equation}
The reversible jump algorithm uses the joint posterior distribution in Equation (\ref{eqn:jointpost}) as the target of a Markov chain Monte Carlo sampler over the state space $\boldsymbol{\Theta}= \bigcup_{k \in {\cal K}} (\{k\}\times {\cal R}^{n_k})$, where the states of the Markov chain are of the form $(k, \boldsymbol{\theta}_k)$,  the
dimension of which can
vary over the state space. 
Accordingly, from the output of a {\it single} Markov chain sampler, the user is able to obtain a full probabilistic description of the posterior probabilities of each model having observed the data, $\Data$, in addition to the posterior distributions of the individual model's parameters.

This article aims to provide an overview of the reversible jump sampler. We outline the sampler's theoretical underpinnings, present some of the most popular and established techniques for enhancing algorithm performance, and discuss the analysis of sampler output. Through the use of several worked examples it is hoped that the reader will gain a broad appreciation of the issues involved in multi-model posterior simulation, and the confidence to implement reversible jump samplers in the course of their own studies. 
Finally, we also briefly outline some recent developments in multi-model    sampling beyond the RJMCMC framework.

\subsection{From Metropolis-Hastings to reversible jump}
\label{sec:MHtoRJ}

The standard formulation of the Metropolis-Hastings algorithm \citep{hastings70} relies on the construction of a time-reversible Markov chain via the {\it detailed balance} condition.  
This condition means that moves from state $\boldsymbol{\theta}$ to $\boldsymbol{\theta'}$ are made as often as  moves from $\boldsymbol{\theta'}$ to $\boldsymbol{\theta}$ with respect to the target density.  This is a simple
way to ensure that the equilibrium distribution of the chain is the desired target distribution.
The extension of the Metropolis-Hastings algorithm to the setting where the dimension of the parameter vector varies is more challenging theoretically, however the resulting algorithm is surprisingly simple
to follow.
%

For the construction of a Markov chain on a general state space $\boldsymbol{\Theta}$ with invariant or stationary distribution $\pi$, the detailed balance condition 
can be written as
\begin{equation}\label{eqn:db1}
\int_{(\boldsymbol{\theta}, \boldsymbol{\theta}') \in {\cal A} \times {\cal B}} \pi(d\boldsymbol{\theta})P(\boldsymbol{\theta}, d\boldsymbol{\theta}') = \int_{(\boldsymbol{\theta}, \boldsymbol{\theta}') \in {\cal A} \times {\cal B}}\pi(d\boldsymbol{\theta}')P(\boldsymbol{\theta}', d\boldsymbol{\theta}) 
\end{equation}
for all Borel sets ${\cal A} \times {\cal B} \subset \boldsymbol{\Theta}$, where $P$ is a general Markov transition kernel \citep[e.g.][]{green01}. 
 
As with the standard Metropolis-Hastings algorithm, Markov chain transitions from a current state $\boldsymbol{\theta} = (k, \boldsymbol{\theta}'_k) \in {\cal A} $ in model ${\cal M}_k$  are realised by first proposing a new state $\boldsymbol{\theta'}=(k', \boldsymbol{\theta}'_{k'}) \in{\cal B}$ in model ${\cal M}_{k'}$  from a proposal distribution $q( \boldsymbol{\theta},  \boldsymbol{\theta'})$. The detailed balance condition 
(\ref{eqn:db1}) is enforced through the acceptance probability, where 
the move to the candidate state $\boldsymbol{\theta'}$ is accepted with probability $\alpha( \boldsymbol{\theta},   \boldsymbol{\theta'})$. If rejected, the chain 
remains at the current state $ \boldsymbol{\theta}$ in model ${\cal M}_k$. 
Under this mechanism \citep{green01,green03},  Equation (\ref{eqn:db1}) becomes
\begin{equation}\label{eqn:db2}
\int_{(\boldsymbol{\theta}, \boldsymbol{\theta'}) \in {\cal A} \times {\cal B}} \pi(\boldsymbol{\theta} | \Data)
q(\boldsymbol{\theta}, \boldsymbol{\theta'})\alpha(\boldsymbol{\theta}, \boldsymbol{\theta'})
d\boldsymbol{\theta}d\boldsymbol{\theta'} = \int_{(\boldsymbol{\theta}, \boldsymbol{\theta'}) \in {\cal A} \times {\cal B}}\pi(\boldsymbol{\theta'}| \Data)q(\boldsymbol{\theta'}, \boldsymbol{\theta}) \alpha(\boldsymbol{\theta'}, \boldsymbol{\theta})d\boldsymbol{\theta}d\boldsymbol{\theta'},
\end{equation}
where the distributions $\pi(\boldsymbol{\theta} | \Data)$
and $\pi(\boldsymbol{\theta'} | \Data)$ are posterior distributions
with respect to model ${\cal M}_k$ and ${\cal M}_{k'}$ respectively.

%
One way to enforce Equation (\ref{eqn:db2}) is by setting the acceptance probability as
\begin{equation}\label{eqn:accept}
\alpha(\boldsymbol{\theta}, \boldsymbol{\theta'}) = \min \left\{ 1, \frac{\pi(\boldsymbol{\theta'}| \Data)q(\boldsymbol{\theta'}, \boldsymbol{\theta})}{\pi(\boldsymbol{\theta} | \Data)
q(\boldsymbol{\theta}, \boldsymbol{\theta'})}\right\},
\end{equation}
where $\alpha(\boldsymbol{\theta'}, \boldsymbol{\theta})$ is similarly defined. This
resembles the usual Metropolis-Hastings acceptance ratio \citep{green95,tierney98}.
It is straightforward to observe that this formulation includes the standard Metropolis-Hastings algorithm as a special case.

Accordingly, a reversible jump sampler with $N$ iterations is commonly constructed as: 
\begin{itemize}
\item[Step 1:] Initialise $k$ and $\boldsymbol{\theta}_k$ at iteration $t=1$.
\item[Step 2:] For iteration $t \geq 1$ perform
\begin{itemize}
\item {\it Within-model move:} with a fixed model $k$, update the parameters $\boldsymbol{\theta}_k$ according to any MCMC updating scheme.
\item {\it Between-models move:} simultaneously update model indicator $k$ and the parameters
$\boldsymbol{\theta}_k$ according to the general reversible proposal/acceptance mechanism (Equation \ref{eqn:accept}).
\end{itemize}
\item[Step 3:] Increment iteration $t=t+1$. If $t<N$, go to Step 2.

\end{itemize}

\subsection{Application areas}\label{sec:app}

Statistical problems in which the number of unknown model parameters is itself unknown are 
extensive,
and as such the reversible jump sampler has been implemented in analyses throughout a wide range of scientific disciplines. Within the statistical literature, these
predominantly concern  Bayesian model determination problems \citep{sisson04,kass+r95}.
Some of the commonly recurring models in this setting are described below.

\begin{description}
\item[Change-point models: ] One of the original applications of the reversible jump sampler was in Bayesian change-point problems, where both the number and location of 
change-points in a system is unknown {\it a priori}. For example, \citet{green95} analysed mining disaster count data 
using a Poisson process with the rate parameter described as a step function with an unknown number and location of steps.
 \citet{fan+b00} applied the reversible jump sampler to model the shape of prehistoric tombs, where the curvature of the dome changes an unknown number of times. Figure \ref{fig:enzyme}(a) shows the plot of depths and radii of one of the tombs from Crete in Greece. The data appear to be piecewise log-linear, with possibly two or three change-points. \citet{boltonh2018} extended the reversible jump sampler for change point detection to also incorporate regime-switching, to infer instruction trace of malware software in a cyber-security setting. \citet{zhaochu2010} developed a model to identify multiple abrupt regime shifts in extreme weather events.  


\begin{figure}[htb]
\begin{center}
\psfrag{stylos data}[l][l]{{\tiny Stylos data}}
\psfrag{Enzyme data}[t][t]{{\tiny Enzyme data}} 
\begin{subfigure}[]{0.45\textwidth}
{
\includegraphics[height=6.75cm,width=7.5cm]{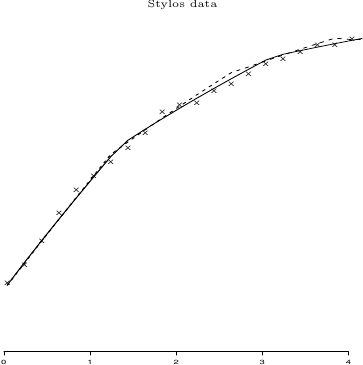}
} 
\end{subfigure}
\begin{subfigure}[]{0.45\textwidth}
{
\includegraphics[height=6.75cm,width=7.5cm]{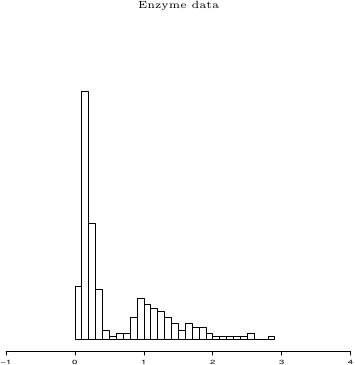}
} 
\end{subfigure}
\caption{\small\label{fig:enzyme} Examples of (a) change-point modelling  and (b) mixture models.  Plot (a): With the Stylos tombs dataset (crosses), a piecewise log-linear curve can be fitted between unknown change-points. Illustrated are 2 (solid line) and 3 (dashed line) change-points. Plot (b): The histogram of the enzymatic activity dataset
suggests clear groupings of metabolizers, although the number of such groupings is not clear.
}
\end{center}
\end{figure}

\item[Finite mixture models: ] 

Mixture models are commonly used where each data observation is generated according to some underlying  categorical mechanism. This mechanism is typically unobserved, so there is uncertainty regarding which component of the resulting mixture distribution each data observation was generated from, in addition to uncertainty over the number of mixture components. A mixture model with $k$ components for the observed data $\Data$ takes the form
\begin{equation}\label{eqn:mixture}
f(\Data|\boldsymbol{\theta}_k) = \sum_{j=1}^k w_j f_j(\Data| \boldsymbol{\phi}_j)
\end{equation}
with $\boldsymbol{\theta_k}=(\boldsymbol{\phi}_1,\ldots,\boldsymbol{\phi}_k,w_1,\ldots,w_{k-1})$,
where $w_j$ is the weight of the $j^{th}$ mixture component $f_{j}$, whose
parameter vector is denoted by $\boldsymbol{\phi}_j$, and where $\sum_{j=1}^kw_j=1$. The number of mixture components, $k$, is also unknown.

Figure \ref{fig:enzyme}(b)  illustrates the distribution of 
enzymatic activity in the blood for 245 individuals. \citet{richardson+g97} analysed these data
using a mixture of Normal densities to identify subgroups of slow or fast metabolizers. The multi-modal nature of the 
data 
suggests the existence of such groups, but the number of distinct groupings is
less clear. Many extensions of mixture component $f_j$ can be found, for example
\citet{NIPS1997_0ed94223} applied the reversible jump to multivariate spherical Gaussian mixtures, and \citet{SALASGONZALEZ2009250} to mixture of $\alpha$-stable distributions.

\item[Variable selection: ] 

The problem of variable selection arises when modelling the relationship between a response variable, $Y$, and 
$p$ potential explanatory variables $x_1,\ldots x_p$. The multi-model setting emerges when attempting to identify the most relevant subsets of predictors, making it a natural candidate for the reversible jump sampler.
For example, 
under a regression model with Normal errors we have
\begin{equation}\label{eqn:varselec}
Y=X_{\gamma}\beta_{\gamma} + \epsilon\qquad\mbox{ with }\qquad \epsilon \sim N(0, \sigma^2I)
\end{equation}
where $\gamma=(\gamma_1,\ldots,\gamma_p)$
is a binary vector indexing the subset of $x_1,\ldots x_p$ to be included in the linear model,  $X_{\gamma}$ is the design matrix whose columns correspond to the indexed subset given by $\gamma$,  and  $\beta_{\gamma} $ is the corresponding subset of regression coefficients. Various extensions to more complex settings have also been proposed. See \citet{nott+l04}  and \cite{fostergo2012} for generalised linear and mixed models,  \citet{lamnisosgs2009} for probit regression, and \citet{newcombeabppr2017} for Weibull regression.

It is well known that regression splines can be estimated within the linear model framework. 
Many authors have successfully explored the use of the reversible jump sampler 
as a method to automate the knot selection process when using a $P$-th order spline model for curve fitting \citep{denison+MS98,dimatteo+gk01}.  Here,  a curve $f$ is estimated by
\begin{equation}
f(x)=\alpha_0 + \sum_{j=1}^P \alpha_jx^j + \sum_{i=1}^k \eta_i(x-\kappa_i)_+^P, \quad x \in [a, b]
\end{equation}
where $z_+=\max(0,z)$ and $\kappa_i, i=1,\ldots,k$, represent the locations of  $k$ knot points \citep{hastie+t90}. Under this representation, 
fitting the curve consists of estimating the unknown number of knots $k$, the knot locations $\kappa_i$ 
and the corresponding regression coefficients $\alpha_j$ and $\eta_i$, for $j=0,\ldots ,P$ and $i=1,\ldots,k$.  
For examples and algorithms in this setting and beyond see e.g. \citet{george+m93}, \citet{smith+k96},  \citet{andrieu+fd00}, \citet{fan2010}.


\item[Bayesian Neural Networks:]
The feed-forward neural network, or multilayer perceptron, can be thought of as a nonlinear regression or classification model in which explanatory variables $x_1,\ldots, x_p$ (or inputs) are related to the response  (or output) variable $Y$. For instance, a very simple model can be written as 
$$
Y= g \left ( w_{00} + \sum_{j=1}^J w_{0j} f\left(w_{j0}+ \sum_{i=1}^p w_{ji}x_i\right) \right ),
$$
where the weights $w_{ji}$ are the strength of connections between the input nodes corresponding to $x_i$ and the $j$th node of the hidden layer. $f$ is the activation function at each of the hidden nodes, and $g$ is the activation function at the output node \citep{titterington2004}. 

Under this setting, we may be interested in models involving different input variables $x_i$, where an inclusion or exclusion of a single $x_i$ may lead to multiple inclusion/exclusion of the weights. Alternatively, we may be interested in models with different structures, where there the number of hidden nodes $J$ may vary \citep{muller1998}.  \citet{holmes1998} and \citet{Berezowski2022} use reversible jump to account for uncertainty in the architecture and depths of the neural network model.

\item[Tree-based models: ]  Motivated by difficulty of designing a well-mixing change-point sampler for latent variable imaging,  \citet{hawkins_transdimensional_2015} introduced a tree-based representation for geophysical images whereby varying tree depths from root to active (or ``leaf'') nodes permits multi-resolution analyses of images in more than one dimension. Furthermore, the mapping from the tree representation to the image space can be specified by any orthogonal basis such as wavelets. Given a tree arrangement $\mc{T}_k$ for a given number of nodes $k$, the conditional prior of the arrangement $p(\mc{T}_k|k)$ has a support that is combinatorial in size. 
\item[Matrix factorisation:] Bayesian interpretation of non-unique factorisation problems is addressed in a factor analysis setting \citep{lopes_bayesian_2004}, and in the more constrained non-negative matrix factorisation setting \citep{zhong_reversible_2009}. The latter addresses the formulation $\bm{X}=\bm{A}\bm{S}+\bm{E}$ where $\bm{A}\in\mc{R}^{N\times M}_+$,  $\bm{S}\in\mc{R}^{M\times T}_+$, and $\bm{E}\in\mc{R}^{N\times T}$, where $M$ is the number of components that varies, and applies an RJMCMC approach for this factorisation in a multiplexed Raman spectra inference example.
\end{description}

The reversible jump algorithm has had a compelling
influence in the statistical and mainstream scientific research literatures,
particularly in 
computationally or
biologically related areas \citep{sisson04}. Accordingly a large number of developmental
and application studies can be found in the signal processing
literature and the related fields of computer vision and image
analysis. Epidemiological and medical studies also feature
strongly.  

This article is structured as follows: In Section \ref{sec:implem} we present discussion on methods for designing between-model moves in the reversible jump sampler,  and Section \ref{sec:improve} reviews approaches to improve sampler performance. Section \ref{sec:converge} details convergence diagnostic tools, followed by discussions on model choice and computing Bayes factors in Section \ref{sec:BF}. In Section \ref{sec:related} we review related multi-model sampling frameworks beyond reversible jump, and in Section \ref{sec:discuss} conclude with discussion on possible future research directions for the field. 

\section{Design of mapping functions and proposal distributions}
\label{sec:implem}


Mapping functions effectively express functional relationships between the parameters of different models.
Good mapping functions will improve reversible jump sampler performance
in terms of between-model acceptance rates and chain mixing. The difficulty is that even in the simpler setting of nested models, good relationships can be hard to define, and in more general settings, parameter vectors between models may not be obviously comparable.
Contrast this to within-model, random-walk Metropolis-Hastings moves on a continuous target density, whereby proposed moves close to the current state can have an arbitrarily large acceptance probability, and proposed moves far from the current state have low acceptance probabilities. Here we discuss some popular strategies for constructing between-model moves.

\subsection{Birth/death and split/merge}

One of the earliest approaches for the construction of proposal moves between different models is achieved via the concept of  ``birth/death" or "split/merge" moves. 
Most simply, under a general Bayesian model determination setting, 
suppose that we are currently in state $(k, \boldsymbol{\theta}_k)$ in model ${\cal M}_k$,  and we wish to propose a move to a state $(k', \boldsymbol{\theta}'_{k'})$ in model ${\cal M}_{k'}$, which is of a higher dimension, so that  $n_{k'} >  n_k $. In order to ``match dimensions'' between the two model states,  a random vector $\boldsymbol{u}$
of length $d_{k\rightarrow k'}=n_{k'} -  n_k$ is generated from a known density $q_{d_{k\rightarrow k'}}(\boldsymbol{u})$. 
The current state $\boldsymbol{\theta}_k$ and the random vector $\boldsymbol{u}$ are then mapped to the new state $\boldsymbol{\theta}'_{k'} = g_{k \rightarrow k'}(\boldsymbol{\theta}_k, \boldsymbol{u})$
through a one-to-one mapping function $g_{k \rightarrow k'} : {\cal R}^{n_k}\times {\cal R}^{d_k} \rightarrow  
{\cal R}^{n_{k'}}$.
 The acceptance probability of this proposal, 
combined with the joint posterior expression of Equation (\ref{eqn:jointpost}) becomes
\begin{equation}\label{eqn:dmatching}
\alpha[(k, \boldsymbol{\theta}_k), (k', \boldsymbol{\theta}'_{k'})] = \min \left\{1, \frac{\pi(k', \boldsymbol{\theta}'_{k'} | \Data)q(k'\rightarrow k)}{\pi(k, \boldsymbol{\theta}_k | \Data)q(k \rightarrow k')q_{d_{k\rightarrow k'}}(\boldsymbol{u})} \left | \frac{\partial g_{k \rightarrow k'}(\boldsymbol{\theta}_k, \boldsymbol{u})}{\partial (\boldsymbol{\theta}_k, \boldsymbol{u})}\right |\right\},
\end{equation}
where $q(k \rightarrow k')$ denotes the probability of proposing a move from model ${\cal M}_k$ to model ${\cal M}_{k'}$, and the final term is the determinant of the Jacobian matrix, often
referred to in the reversible jump literature simply as the Jacobian. This term arises through the change of variables via the function $g_{k\rightarrow k'}$, which is required when used with respect to the integral Equation  (\ref{eqn:db2}).
Note that the normalisation constant in Equation (\ref{eqn:jointpost}) is not needed to evaluate the above ratio. The reverse move proposal, from model ${\cal M}_{k'}$ to ${\cal M}_k$ is made deterministically in this setting, and
is accepted with probability 
$$
\alpha[(k', \boldsymbol{\theta}'_{k'}), (k, \boldsymbol{\theta}_{k})] =\alpha[(k, \boldsymbol{\theta}_k), (k', \boldsymbol{\theta}'_{k'})]^{-1}.
$$
More generally, we can relax the condition on the length of the vector $\boldsymbol{u}$
by allowing $d_{k\rightarrow k'}\geq n_{k'}-n_k$. In this case, non-deterministic reverse moves can be made by generating a $d_{k'\rightarrow k}$-dimensional random vector $\boldsymbol{u}'~\sim q_{d_{k'\rightarrow k}}(\boldsymbol{u}')$,
such that the dimension matching condition, $n_k+d_{k\rightarrow k'}=n_{k'}+d_{k'\rightarrow k}$, is satisfied.
Then a reverse mapping is given by
$\boldsymbol{\theta}_k=g_{k'\rightarrow k}(\boldsymbol{\theta}'_{k'}, \boldsymbol{u}')$, such
that $\boldsymbol{\theta}_k=g_{k'\rightarrow k}(g_{k \rightarrow k'}(\boldsymbol{\theta}_k, \boldsymbol{u}),\boldsymbol{u}')$ and $\boldsymbol{\theta}'_{k'}=g_{k\rightarrow k'}(g_{k' \rightarrow k}(\boldsymbol{\theta}'_{k'}, \boldsymbol{u}'),\boldsymbol{u})$. 
%
The corresponding acceptance probability to Equation (\ref{eqn:dmatching}) then becomes
\begin{equation}\label{eqn:dmatching2}
\alpha[(k, \boldsymbol{\theta}_k), (k', \boldsymbol{\theta}'_{k'})] = \min \left\{1, \frac{\pi(k', \boldsymbol{\theta}'_{k'} | \Data)q(k'\rightarrow k)q_{d_{k'\rightarrow k}}(\boldsymbol{u}')}{\pi(k, \boldsymbol{\theta}_k | \Data)q(k \rightarrow k')q_{d_{k\rightarrow k'}}(\boldsymbol{u})} \left | \frac{\partial g_{k \rightarrow k'}(\boldsymbol{\theta}_k, \boldsymbol{u})}{\partial (\boldsymbol{\theta}_k, \boldsymbol{u})}\right |\right\}.
\end{equation}

\noindent{\bf Example:  Simple birth/death and split/merge}\\
Consider the illustrative example given in \citet{green95} and
\citet{brooks98}. Suppose that model ${\cal M}_1$ has states $(k=1, \boldsymbol{\theta}_1 \in {\cal R}^1)$ and model ${\cal M}_2$ has states  $(k=2, \boldsymbol{\theta}_2 \in {\cal R}^2)$. Let $(1, \theta^*)$ denote the current state in ${\cal M}_1$ and $(2, (\theta^{(1)}, \theta^{(2)})$) denote the proposed state in ${\cal M}_2$. Under dimension matching with a simple split/merge move, we might generate a random scalar $u$, and 
let $\theta^{(1)} = \theta^*+u$ and $\theta^{(2)} = \theta^* - u$, with the reverse move given deterministically by $\theta^* = \frac{1}{2}(\theta^{(1)}+\theta^{(2)})$.
For the same setup but with a simple birth/death move, we might specify $\theta^{(1)} = \theta^*$ and $\theta^{(2)} = \theta^*+u$, with the reverse move given deterministically by $\theta^* = \theta^{(1)}$.

\noindent{\bf Example: Moment matching in a finite mixture of univariate Normals}\\
Under the finite mixture of univariate
Normals model, 
the observed data, $\Data$, has
density given by Equation (\ref{eqn:mixture}), where the $j$-th mixture component
$f_j(\Data|\boldsymbol{\phi}_j)=
\phi(\Data|\mu_j,\sigma_j)$ is the $N(\mu_j,\sigma_j)$ density.
For between-model moves,
\citet{richardson+g97} 
implement a  split (one component into two) and merge (two components into one) strategy which 
satisfies the dimension matching requirement. \cite[See][for an alternative approach]{dellaportas+p06}.

When two Normal components $j_1$ and $j_2$ are merged into one, $j^*$,
\citet{richardson+g97} propose a deterministic mapping which maintains the
$0^{\mbox{\tiny th}}, 1^{\mbox{\tiny st}}$ and $2^{\mbox{\tiny
nd}}$ moments:
\begin{equation}\label{eqn:MMmerge}\begin{array}{lll}
w_{j^*} & = & w_{j_1}+w_{j_2}\\
w_{j^*}\mu_{j^*} & = & w_{j_1}\mu_{j_1}+w_{j_2}\mu_{j_2}\\
w_{j^*}(\mu^2_{j^*}+\sigma^2_{j^*}) & = & w_{j_1}(\mu^2_{j_1}+\sigma^2_{j_1})+w_{j_2}(\mu^2_{j_2}+\sigma^2_{j_2}).
\end{array}
\end{equation}
The split move is proposed as
\begin{equation}\label{eqn:MMsplit}\begin{array}{lll}
w_{j_1} = &w_{j^*}*u_1,&  w_{j_2} = w_{j^*}*(1-u_1)\\
\mu_{j_1}&=&\mu_{j^*} - u_2\sigma_{j^*}\sqrt{\frac{w_{j_2}}{w_{j_1}}}\\
\mu_{j_2}&=&\mu_{j^*} + u_2\sigma_{j^*}\sqrt{\frac{w_{j_1}}{w_{j_2}}}\\
 \sigma^2_{j_1}&=&u_3(1-u_2^2)\sigma^2_{j^*}\frac{w_{j^*}}{w_{j_1}}\\
 \sigma^2_{j_2}&=&(1-u_3)(1-u_2^2)\sigma^2_{j^*}\frac{w_{j^*}}{w_{j_2}},
\end{array}
\end{equation}
with the random scalars $u_1, u_2\sim\mbox{Beta}(2,2)$ and $u_3\sim\mbox{Beta}(1,1)$.
In this manner, dimension matching is satisfied, and the acceptance probability for the split move is calculated according to Equation (\ref{eqn:dmatching}), with the acceptance probability of the reverse merge move given by the reciprocal of this value.



While the ideas behind dimension matching are conceptually simple, their implementation
is complicated by the arbitrariness of the mapping function $g_{k \rightarrow k'}$ and the proposal distributions, $q_{d_{k\rightarrow k'}}(\boldsymbol{u})$, for the random vectors $\boldsymbol{u}$.

\subsection{Centering and order methods}

The concept of  ``local" moves, akin to that of random-walk Metropolis-Hastings in fixed dimensions,  may be partially translated on to model space ($k\in{\mathcal K}$): proposals from $\boldsymbol{\theta}_k$  in model ${\mathcal M}_k$ to $\boldsymbol{\theta}'_{k'}$  in model ${\mathcal M}_{k'}$ will tend to have larger acceptance probabilities if their likelihood values are similar i.e.~$L(\Data| k,\boldsymbol{\theta}_k)\approx L(\Data| k',\boldsymbol{\theta}'_{k'})$.

\citet{brooks+gr03} introduce a class of methods
to achieve the automatic scaling of
the proposal density, $q_{d_{k\rightarrow k'}}(\boldsymbol{u})$, 
based on the concept of the ``local'' move proposal distributions.
%
%
Under this scheme, 
it is assumed that local mapping functions $g_{k\rightarrow k'}$ are known.
 For a proposed move from $(k, \boldsymbol{\theta_k})$ in
${\cal M}_k$ to model ${\cal M}_{k'}$, 
the random vector ``centering point''
$c_{k\rightarrow k'}(\boldsymbol{\theta}_k)=g_{k\rightarrow k'}(\boldsymbol{\theta}_k,\boldsymbol{u})$, is defined such
that for some particular choice of proposal vector
$\boldsymbol{u}$, the current and proposed states are identical in
terms of likelihood contribution. 
 %

Given the centering constraint on $\boldsymbol{u}$, if the scaling parameter in the proposal $q_{d_{k\rightarrow k'}}(\boldsymbol{u})$ is a scalar, then the  $0^{th}$-order method \citep{brooks+gr03} proposes to choose this scaling parameter such that the acceptance probability $\alpha[(k, \boldsymbol{\theta}_k),(k', c_{k\rightarrow k'}(\boldsymbol{\theta}_k))]$ of a move to the centering point $c_{k\rightarrow k'}(\boldsymbol{\theta}_k)$ in model $\mathcal{M}_{k'}$ 
 is exactly one. The argument is then that move proposals close to  $c_{k\rightarrow k'}(\boldsymbol{\theta}_k)$ will also have a large acceptance probability.

For proposal distributions, $q_{d_{k\rightarrow k'}}(\boldsymbol{u})$, with additional degrees of freedom,
a similar method 
 based on a series of $n^{th}$-order conditions (for $n\geq 1$), requires that for
the proposed move, the $n^{th}$ derivative (with respect to $\boldsymbol{u}$) of the acceptance
probability equals the zero vector at the centering point $c_{k\rightarrow k'}(\boldsymbol{\theta}_k)$:
\begin{equation}\label{eqn:nth_order} 
\nabla^n\alpha[(k, \boldsymbol{\theta}_k),(k', c_{k\rightarrow k'}(\boldsymbol{\theta}_k))]=\boldsymbol{0}.
\end{equation}
That is, the $m$ unknown parameters in the proposal
distribution $q_{d_{k\rightarrow k'}}(\boldsymbol{u})$ are determined by
solving the $m$ simultaneous equations given
by (\ref{eqn:nth_order}) with $n=1, \ldots, m$.
The idea behind the $n^{th}$-order method is that the concept of closeness to the centering point under the $0^{th}$-order method is relaxed.
By enforcing zero derivatives of $\alpha[(k, \boldsymbol{\theta}_k),(k', c_{k\rightarrow k'}(\boldsymbol{\theta}_k))]$, the acceptance probability will become flatter around $c_{k\rightarrow k'}(\boldsymbol{\theta}_k)$. Accordingly this allows proposals  further away from the centering point to still be accepted with a reasonably high probability.
%
This will ultimately induce improved chain mixing.

One caveat with the centering schemes is that they require specification of the between
model mapping function $g_{k\rightarrow k'}$, although these methods
compensate for poor choices of mapping functions by
selecting the best set of parameters for the given mapping.
\citet{ehlers+b08} suggest the posterior conditional 
distribution $\pi(k', \boldsymbol{u}| \boldsymbol{\theta}_k)$ as the proposal for
the random vector $\boldsymbol{u}$, side-stepping the need to construct a mapping function. In this case, the full conditionals must either be known, or need to be approximated. 

\noindent{\bf Example: The $0^{th}$-order method for an autoregressive model}\\
 \citet{brooks+gr03} considers the AR model for temporally dependent observations $x_1,\ldots x_T$,  with unknown order $k$ 
\begin{equation*}
	\label{eqn:timeseries}
	X_t = \sum_{\tau=1}^k a_{\tau}X_{t-\tau} +\epsilon_t\qquad\mbox{ with }\qquad t=k+1, \ldots,T,
\end{equation*}
assuming Gaussian noise $\epsilon_t \sim N(0,\sigma_{\epsilon}^2)$ and a Uniform prior on  $k$ where $k=1,2,\ldots k_{max}$. Within each model ${\cal M}_k$,
independent $N(0,\sigma_a^2)$ priors are adopted for the AR coefficients $a_{\tau}, \tau=1,\ldots, k$, with an Inverse Gamma prior for $\sigma^2_{\epsilon}$.
Suppose moves are made 
from model ${\cal M}_k$  to model ${\cal M}_{k'}$ such that
$k'=k+1$. The move from 
$\boldsymbol{\theta}_k$ to $\boldsymbol{\theta}'_{k'}$ is achieved by generating a random scalar $u \sim q(u)=N(0, 1)$, and defining
the mapping function as $\boldsymbol{\theta}'_{k'} = g_{k \rightarrow k'}(\boldsymbol{\theta}_k, u) =  (\boldsymbol{\theta}_k,  \sigma u)$. The centering point  $c_{k\rightarrow k'}(\boldsymbol{\theta}_k)$ then occurs at the point $u=0$, or $\boldsymbol{\theta}'_{k'} = (\boldsymbol{\theta}_k, 0)$.

Under the mapping $g_{k \rightarrow k'}$, the Jacobian is  $\sigma$, and the acceptance probability  (Equation \ref{eqn:dmatching}) for the move from $(k,\boldsymbol{\theta}_k)$ to 
$(k',c_{k\rightarrow k'}(\boldsymbol{\theta}_{k}))$ is given by $\alpha[(k, \boldsymbol{\theta}_k), (k', (\boldsymbol{\theta}_k, 0))]=\min(1,A)$ where
\[
A=\frac{\pi(k', (\boldsymbol{\theta}_{k},0) | \Data)q(k'\rightarrow k)\sigma}{\pi(k, \boldsymbol{\theta}_k | \Data)q(k \rightarrow k')q(0)}
 = 
\frac{(2\pi\sigma_a^2)^{-1/2}q(k'\rightarrow k)\sigma}{q(k \rightarrow k') (2\pi)^{-1/2}}.
\]
Note that since the likelihoods are equal at the centering point,
and the priors common to both models cancel in the posterior ratio, $A$ is only a function of
the prior density for the parameter $a_{k+1}$ evaluated at 0, the proposal distributions and the Jacobian. Hence we solve $A=1$
to obtain
$$\sigma^2=\sigma^2_a\left(\frac{q(k\rightarrow k')}{q(k' \rightarrow k)}\right)^2.$$
Thus in this case, the proposal variance is not model parameter ($\boldsymbol{\theta}_k$) or data ($\Data$) dependent. 
It depends only on the prior variance, $\sigma_a$, and the model states, $k,k'$.

\noindent{\bf Example: The second-order method for moment matching }\\
Consider the moment matching in a finite mixture of univariate Normals example of Section \ref{sec:app}. The mapping functions $g_{k'\rightarrow k}$ and $g_{k\rightarrow k'}$ are respectively given by Equations (\ref{eqn:MMmerge}) and (\ref{eqn:MMsplit}), with 
the random numbers $u_1, u_2$ and $u_3$  drawn from independent Beta distributions with unknown parameter values, so that $q_{p_i,q_i}(u_i)$: $u_i\sim\mbox{Beta}(p_i,q_i)$, $i=1,2,3.$  

Consider the split move, Equation (\ref{eqn:MMsplit}). To apply the second order method of \citet{brooks+gr03}, we first locate a centering
point,
$c_{k\rightarrow k'}(\boldsymbol{\theta}_k)$, achieved by setting  $u_1=1$, $u_2=0$ and $u_3\equiv u_1 =1$ by inspection. 
Hence, at the centering point, 
 the two new
(split) components $j_1$ and $j_2$ will have the same location and scale as the $j^*$ component, with new weights
$w_{j_1}=w_{j^*}$ and $w_{j_2}=0$ and all observations allocated to component $j_1$. Accordingly this will produce identical likelihood contributions.
Note that to obtain equal variances for the split proposal, substitute the expressions for $w_{j_1}$ and $w_{j_2}$ into those for $\sigma^2_{j_1}=\sigma^2_{j_2}$.

Following \citet{richardson+g97}, the acceptance probability of the split move evaluated at the centering point
is then proportional (with respect to $\boldsymbol{u}$) to
\begin{equation}\label{eqn:MMacc}\begin{array}{lll}
&&\log A[(k, \boldsymbol{\theta}_k),(k', c_{k\rightarrow k'}(\boldsymbol{\theta}_k))] \propto\\
&&\qquad l_{j_1}\log(w_{j_1}) + l_{j_2}\log(w_{j_2})-\frac{l_{j_1}}{2}\log(\sigma^2_{j_1})-\frac{l_{j_2}}{2}\log(\sigma^2_{j_2})
-\frac{1}{2\sigma^2_{j_1}}\sum_{l=1}^{l_{j_1}}(y_l-\mu_{j_1})^2\\
&&\qquad-\frac{1}{2\sigma^2_{j_2}}\sum_{l=1}^{l_{j_2}}(y_l-\mu_{j_2})^2+(\delta-1+l_{j_1})\log(w_{j_1}) + (\delta-1+l_{j_2})\log(w_{j_2})\\
&&\qquad -\{\frac{1}{2}\kappa[(\mu_{j_1}-\xi)^2+(\mu_{j_2}-\xi)^2]\} -(\alpha+1)\log(\sigma^2_{j_1}\sigma^2_{j_2}) -\beta(\sigma^{-2}_{j_1}+\sigma^{-2}_{j_2})\\
&&\qquad-\log [q_{p_1,q_1}(u_1)]-\log [q_{p_2,q_2}(u_2)]-\log [q_{p_3,q_3}(u_3)]+\log( |\mu_{j_1}-\mu_{j_2}|)\\
&&\qquad +\log (\sigma^2_{j_1}) +\log (\sigma^2_{j_2})-\log(u_2)-\log(1-u_2^2)-\log(u_3)-\log(1-u_3),
\end{array}
\end{equation}
where 
$l_{j_1}$ and $l_{j_2}$ 
respectively denote the number of observations allocated to components $j_1$ and $j_2$,
and where  $\delta,\alpha,\beta,\xi$ and $\kappa$ are hyperparameters as defined by
\citet{richardson+g97}.
 
Thus, for example, to obtain the proposal parameter values $p_1$ and $q_1$ for $u_1$, 
we solve the first- and second-order derivatives of the acceptance probability (\ref{eqn:MMacc}) with respect to $u_1$. This yields
$$
\frac{\partial \log \alpha[(k, \boldsymbol{\theta}_k),(k', c_{k\rightarrow k'}(\boldsymbol{\theta}_k))]}{\partial u_1}
=\frac{\delta+2l_{j_1}-p_1}{u_1}+\frac{q_1-\delta-2l_{j_2}}{(1-u_1)}
$$
$$
\frac{\partial^2 \log \alpha[(k, \boldsymbol{\theta}_k),(k', c_{k\rightarrow k'}(\boldsymbol{\theta}_k))]}{\partial u^2_1}
=-\frac{\delta+2l_{j_1}-p_1}{u^2_1}+\frac{q_1-\delta-2l_{j_2}}{(1-u_1)^2}.
$$
Equating these to zero and solving for $p_1$ and $q_1$ at the centering points (with $l_{j_1}=l_{j^*}$ and $l_{j_2}=0$) gives $p_1=\delta+2l_{j^*}$ and  $q_1=\delta$. Thus the parameter $p_1$ depends on the number of observations allocated to the component being split. Similar calculations to the above give solutions for $p_2, q_2, p_3$ and $q_3$.

\subsection{Generic samplers}

The problem of efficiently constructing between-model mapping
templates, $g_{k\rightarrow k'}$, with associated random vector proposal densities, $q_{d_{k\rightarrow k'}}$,
may be approached from an alternative perspective.
Rather than relying on a user-specified mapping,
one strategy would be to move towards a more generic proposal mechanism altogether.  A clear benefit of generic between-model moves is that they may be equally be implemented for non-nested models.

\citet{green03} proposed
 a reversible jump analogy of the random-walk Metropolis sampler of
\citet{roberts03}. 
Suppose that estimates of
the first and second order moments of $\boldsymbol{\theta}_k$ are available, for each of a small number of models, $k\in{\mathcal K}$,
denoted  by $\boldsymbol{\mu}_k$ and $\boldsymbol{B}_k\boldsymbol{B}^{\top}_k$ respectively, where $\boldsymbol{B}_k$ is an $n_k \times n_k$ matrix. In proposing
a move from $(k, \boldsymbol{\theta}_k)$ to model ${\cal M}_{k'}$, a new parameter
vector is proposed by
\begin{equation*}
\boldsymbol{\theta}'_{k'} = \left\{\begin{array}{ll} \boldsymbol{\mu}_{k'} +
\boldsymbol{B}_{k'}\left[\boldsymbol{R}\boldsymbol{B}^{-1}_{k}(\boldsymbol{\theta}_k-\boldsymbol{\mu_k)}\right]^{n_{k'}}_1 &
\mbox{if } n_{k'}<n_k\\
\boldsymbol{\mu}_{k'}+\boldsymbol{B}_{k'}\boldsymbol{RB}^{-1}_k(\boldsymbol{\theta}_k-\boldsymbol{\mu}_k) & \mbox{if }n_{k'}=n_k\\
\boldsymbol{\mu}_{k'}+\boldsymbol{B}_{k'}\boldsymbol{R}\left(\begin{array}{c}\boldsymbol{B}^{-1}_k(\boldsymbol{\theta}_k-\boldsymbol{\mu}_k)\\
\boldsymbol{u}\end{array}\right) & \mbox{if }n_{k'}>n_k
\end{array}\right.
\end{equation*}
where  $[\,\cdot\,]_1^m$
denotes the first $m$ components of a vector, $\boldsymbol{R}$ is an
orthogonal matrix of order
$\max\{n_k,n_{k'}\}$, and $\boldsymbol{u}\sim q_{n_{k'}-n_{k}}(\boldsymbol{u})$ is an
$(n_{k'}-n_k)$-dimensional random vector (only utilised if $n_{k'}>n_k$, or when calculating the acceptance probability of the reverse move from model ${\mathcal M}_{k'}$ to model ${\mathcal M}_k$ if $n_{k'}<n_k$). If $n_{k'} \leq n_k$, then the proposal $\boldsymbol{\theta}'_{k'}$ is deterministic and the Jacobian 
is trivially calculated.
Hence the acceptance probability is given by
\begin{equation*}
\alpha[(k, \boldsymbol{\theta}_k),(k', \boldsymbol{\theta}'_{k'})] =
\frac{\pi(k', \boldsymbol{\theta}'_{k'}| \Data)}{\pi(k, \boldsymbol{\theta}_k|\Data)}
\frac{q(k'\rightarrow k)}{q(k\rightarrow k')} \frac{|\boldsymbol{B}_{k'}|}{|\boldsymbol{B}_k|}
\times\left\{\begin{array}{ll} q_{n_{k'}-n_k}(\boldsymbol{u}) & \mbox{for } n_{k'}<n_k\\
1 & \mbox{for }n_{k'}=n_k\\
1/q_{n_{k'}-n_k}(\boldsymbol{u}) & \mbox{for } n_{k'}>n_k
\end{array}\right. .
\end{equation*}
Accordingly, if the model-specific densities $\pi(k, \boldsymbol{\theta}_k|\Data)$
are uni-modal with first and second order moments given by $\boldsymbol{\mu}_k$ and
$\boldsymbol{B}_k\boldsymbol{B}^{\top}_k$, then high between-model acceptance probabilities
may be achieved. 

With a similar motivation to the above,  \citet{papathomas+dv09} propose the multivariate Normal  
as proposal distribution for $\boldsymbol{\theta}'_{k'}$ in the context of linear regression models, 
so that
$\boldsymbol{\theta}'_{k'}\sim N(\boldsymbol{\mu}_{k'|\boldsymbol{\theta}_k}, \boldsymbol{\Sigma}_{k'|\boldsymbol{\theta}_k})$.
The authors derive
estimates for the mean $\boldsymbol{\mu}_{k'|\boldsymbol{\theta}_k}$ and covariance $\boldsymbol{\Sigma}_{k'|\boldsymbol{\theta}_k}$ such that
the proposed values for $\boldsymbol{\theta}'_{k'}$ will on average produce similar
conditional posterior values under model ${\mathcal M}_{k'}$ as the vector $\boldsymbol{\theta}_{k}$ under model ${\mathcal M}_k$. 
The method is theoretically
justified for Normal linear models, but can be applied to non-Normal models when transformation
of data to Normality is available.
\citet{green03}, \citet{godsill03},  \citet{hastie04}, and \citep{farr2015efficient} discuss a number of modifications to
the generic framework approach, including improving efficiency and relaxing the requirement of unimodal densities $\pi(k, \boldsymbol{\theta}_k|\Data)$ to realise high between-model acceptance rates. Naturally, for all Normal-based approximations, the required knowledge of first and second order moments of each model density will restrict the applicability of these approaches to moderate numbers of candidate models if these require estimation (e.g.~via pilot chains). For proposals that use kD-tree approximations to the model densities \citep{farr2015efficient} this restriction is less apparent with a small trade-off in high-dimensionality performance.

%

\begin{figure}
    \centering
    \includegraphics[width=\textwidth]{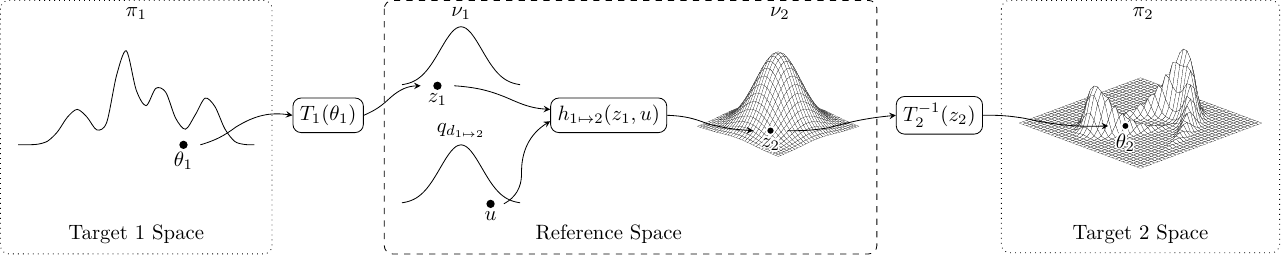}
    \caption{\small A transport-based end-to-end bijective mapping starting a point $\bm{\theta}_1$ in a 1D target space (\textit{left}) to a point $\bm{\theta}_2$ in a 2D target space (\textit{right}). The first transport maps $\bm{\theta}_1$ to the point $\bm{z}_1=T_1(\bm{\theta}_1)$ in the associated reference space, and then the transdimensional bijective transformation $h_{1\mapsto 2}$ augments $\bm{z}_1$ with auxiliary variable $\bm{u}$ in a vector concatenation, resulting in point $\bm{z}_2=h_{1\mapsto 2}(\bm{z}_1,\bm{u})$ in the 2D reference space. Then, $\bm{z}_2$ is mapped to $\bm{\theta}_2$ in the 2D target space via the inverse transport $\bm{\theta}_2=T_2^{-1}(\bm{z}_2)$. }
    \label{fig:trjp}
\end{figure}

A generalised approach to proposal distribution design in MCMC methods when the target distributions 
$\pi(\bm{\theta}_k|k, \Data)$ are strongly non-Normal is to condition via a \emph{transport} \citep{parno_transport_2018}. Deep neural network based \emph{normalising flows} \citep{rezende_variational_2015,papamakarios2021normalizing} perform demonstrably well in the approximation of transports and have been shown to be useful when trained adaptively during MCMC burn-in rather than requiring pilot runs \citep{gabrie_adaptive_2022}.
\citet{davies_transport_2023} generalise generic RJMCMC proposals using such a transport-based approach where the distributions of interest are the conditional targets
$\pi_k$ 
with density functions $\pi(\theta_k|k,\Data)$. In this context, a transport $\bm{z}_k=T_k(\bm{\theta}_k)$, $\bm{z}_k\in\mc{Z}_k$, is a bijective transform of samples from $\pi_k$ to a chosen reference distribution $\nu_k$, and comprises the pushforward $\nu_k=T_k\#\pi_k$. Defining the density of $\nu_k$ on the support of $\mc{Z}_k\subseteq\mc{R}^{n_k}$, this relationship can be expressed using the change of variables 
\[
\pi(\bm{\theta}_k|k, \Data) = \nu_k(T_k(\bm{\theta}_k))\left|\dfrac{\partial T_k(\bm{\theta}_k)}{\partial\bm{\theta}_k}\right|.
\]
The mechanism for the RJMCMC proposal is to allow the chain to jump between reference distributions $\nu_{k}$ to $\nu_{k'}$ instead of directly between the respective conditional targets $\pi_k$ and $\pi_{k'}$. 
This is achieved by first choosing a univariate base distribution $\nu$ (which is typically a standard Normal) with density function $\nu(z)$, $z\in\mc{Z}_1$ and then defining all reference distributions using the formulation
\[\nu_k=\otimes_{n_k}\nu=\underbrace{\nu\otimes\dots\otimes\nu}_{n_k \text{times}}\,\,\text{for each }\,k,
\]
where the respective density of each $\nu_k$ has the form $\nu_k(\bm{z}_k)$, $\bm{z}_k\in\mc{Z}_k$. In essence, this construction ensures that each component of $\bm{z}_k$ is i.i.d.~on $\nu$. A crucial property that is exploited in this construction is that the auxiliary variables required for dimension matching are also defined to be i.i.d.~on $\nu$, that is for auxiliary dimension $d_{k\mapsto k'}$ we have $\bm{u}\sim\otimes_{d_{k\mapsto k'}}\nu$. Next, dimension matching is achieved by defining pairwise volume-preserving transformations $\bm{z}_{k'},\bm{u}'=h_{k\mapsto k'}(\bm{z}_{k'},\bm{u}')$ between points in the supports of respective reference distributions $\nu_k$, $\nu_{k'}$, a simple construction being the vector concatenation $\bm{z}_{k'}\gets [\bm{z}_{k};\, \bm{u}]$ when $n_{k'}>n_k$.
Figure \ref{fig:trjp} depicts an example of the bijective mapping of parameters and auxiliary variables from a 1D target $\pi_1$ to a 2D target $\pi_2$. The transports ensure points distributed according to each target are mapped to points in the respective reference spaces and distributed according to the chosen reference distributions. 
The general case for a mapping between points in the supports of $\pi_k$ and $\pi_{k'}$ is the composition 
\[
(\bm{\theta}_{k'}',\bm{u}')=g_{k\mapsto k'}(\bm{\theta}_k,\bm{u})= T_{k'}^{-1}(h_{k\mapsto k'}(T_k(\bm{\theta}_k),\bm{u})),
\] with Jacobian determinant
\[
\left|\dfrac{\partial g_{k\mapsto k'}(\bm{\theta}_k,\bm{u})}{\partial (\bm{\theta}_k,\bm{u})}\right|
=
\left|\dfrac{\partial T_{k}(\bm{\theta}_{k})}{\partial \bm{\theta}_{k}}\right|
\left|\dfrac{\partial h_{k\mapsto k'}(T_k(\bm{\theta}_{k}),\bm{u})}{\partial (T_k(\bm{\theta}_{k}),\bm{u})}\right|
\left|\dfrac{\partial T_{k'}(\bm{\theta}_{k'}')}{\partial \bm{\theta}_{k'}'}\right|^{-1}.
\] 
Since the pairwise construction of $h_{k\mapsto k'}$ is largely trivial and can be defined for all pairs of reference distributions due to the property that $\bm{z}$ and $\bm{u}$ are i.i.d., jumps between any two models exist by default, allowing global and independent exploration of the model space. When $h_{k\mapsto k'}$ is volume preserving, i.e.~$|\partial h_{k\mapsto k'}(\bm{z}_k,\bm{u})/\partial (\bm{z}_k,\bm{u})|=1$ for $\bm{z}_k=T_k(\bm{\theta}_k)$, the acceptance probability of such a generic transport RJMCMC proposal is
\begin{equation*}
\alpha[(k, \boldsymbol{\theta}_k),(k', \boldsymbol{\theta}'_{k'})] 
=
\frac
{\pi(k', \boldsymbol{\theta}'_{k'}| \Data)}
{\pi(k, \boldsymbol{\theta}_k|\Data)}
\frac
{q(k'\rightarrow k)}
{q(k\rightarrow k')} 
\left|\dfrac{\partial T_{k}(\bm{\theta}_{k})}{\partial \bm{\theta}_{k}}\right|
\left|\dfrac{\partial T_{k'}(\bm{\theta}_{k'}')}{\partial \bm{\theta}_{k'}'}\right|^{-1}.
\end{equation*}

\citet{fan+ps08} propose to construct between-model proposals based on estimating conditional marginal densities.
Suppose that  it is reasonable to assume some structural similarities between the parameters
$\boldsymbol{\theta}_{k}$ and $\boldsymbol{\theta}'_{k'}$ of models ${\mathcal M}_k$ and
${\mathcal M}_{k'}$ respectively. Let $c$ indicate the subset of the vectors $\boldsymbol{\theta}_{k}=(\boldsymbol{\theta}_{k}^c, \boldsymbol{\theta}_{k}^{-c})$ 
and $\boldsymbol{\theta}'_{k'}=(\boldsymbol{\theta}_{k'}'^c, %
\boldsymbol{\theta}_{k'}'^{-c})$ which 
can be kept constant between models,  so that  $\boldsymbol{\theta}_{k'}'^c=\boldsymbol{\theta}_k^c$.
The remaining $r$-dimensional vector $\boldsymbol{\theta}_{k'}'^{-c}$ 
is then sampled from an estimate of the factorisation of the conditional posterior of $\boldsymbol{\theta}_{k'}'^{-c}=(\theta^1_{k'},\ldots,\theta^r_{k'})$ under model ${\mathcal M}_{k'}$:
$$
\pi(\boldsymbol{\theta}^{-c}_{k'}|\boldsymbol{\theta}^c_{k'},\Data)
\approx\hat{\pi}_1(\theta^1_{k'}|
\theta_{k'}'^2,\ldots,\theta_{k'}'^{r},\boldsymbol{\theta}_{k'}'^c,\Data)\ldots
\hat{\pi}_{r-1}(\theta^{r-1}_{k'}|\theta_{k'}'^r,\boldsymbol{\theta}_{k'}'^{c},\Data)\hat{\pi}_{r}(\theta_{k'}'^r|\boldsymbol{\theta}_{k'}'^{c},\Data).
$$
The proposal $\boldsymbol{\theta}^{-c}_{k'}$ is drawn by first estimating $\hat{\pi}_r(\theta_{k'}'^r|\boldsymbol{\theta}^{c}_{k'},\Data)$ and sampling $\theta^r_{k'}$, and by then estimating $\hat{\pi}_{r-1}(\theta^{r-1}_{k'}|\theta_{k'}'^r,\boldsymbol{\theta}^{c}_{k'},\Data)$ and sampling $\theta^{r-1}_{k'}$,  conditioning on the previously sampled point, $\theta^r_{k'}$, and so on.
\citet{fan+ps08} construct the conditional marginal densities by using partial derivatives of the joint
density, $\pi(k', \boldsymbol{\theta}'_{k'}|\Data)$, to provide gradient information within a marginal density estimator.  As the conditional marginal density estimators 
are constructed using a combination of samples from the prior distribution and gridded values, they can be computationally expensive to construct, particularly if high-dimensional moves are attempted e.g.~$\boldsymbol{\theta}^{-c}_{k'}=\boldsymbol{\theta}'_{k'}$. However, this approach can  be efficient, and also adapts to the current state of the sampler.

\section{Schemes to improve sampler performance}\label{sec:improve}

\subsection{Marginalisation and augmentation}

Depending on the aim or the complexity of a
multi-model analysis, it may be that use of reversible jump MCMC would be somewhat heavy-handed, when 
reduced- or fixed-dimensional samplers may be substituted. 
In some Bayesian model selection settings, between-model moves can be greatly 
simplified or even avoided if one is prepared to make certain prior assumptions, 
such as conjugacy or objective prior specifications.
In such cases, it may be possible to analytically integrate out some or all of the parameters
$\boldsymbol{\theta}_k$ in the posterior distribution (\ref{eqn:jointpost}), reducing the sampler either
to fixed dimensions, e.g~ on model space $k\in{\mathcal K}$ only, or to a lower-dimensional set of model and
parameter space \citep{tadessesv05,dimatteo+gk01,berger+p01, drovandiphm2014, persing2015simulation}. 
In lower dimensions, the reversible jump sampler is often easier to implement, as the problems associated with mapping function specification are conceptually simpler to resolve.

\noindent{\bf Example: Marginalisation in variable selection}\\
In Bayesian variable selection for Normal linear models (Equation \ref{eqn:varselec}),
the vector $\gamma= (\gamma_1, \ldots, \gamma_p)$ is treated as an auxiliary (model indicator) variable, where
\begin{equation*}
\gamma_i =\left\{\begin{array}{ll}
1 &\quad \mbox{if predictor } x_i \mbox{ is included in the regression} \\
0 &\quad \mbox{otherwise}.
\end{array}\right.
\end{equation*}
Under certain prior specifications for the regression coefficients $\beta$ and error variance $\sigma^2$, the $\beta$ coefficients can be analytically integrated out of the posterior. A Gibbs sampler directly on model space is then available for  $\gamma$ \citep{george+m93,smith+k96,nott+g04, yang2016computational,zhou2022dimension}.

\noindent{\bf Example: Marginalisation in finite mixture of multivariate Normal models}\\
Within the context of clustering, the parameters of the Normal components are usually not of interest. \citet{tadessesv05} demonstrate that by choosing appropriate prior distributions, the parameters of the Normal  components can be analytically integrated out of the posterior. The reversible jump sampler may then run on a much reduced parameter space, which is  simpler and more efficient.

In a general setting, \citet{brooks+gr03} proposed a class of models based on
augmenting the state space of the target posterior with an auxiliary set of
state-dependent variables, $\boldsymbol{v}_k$, so that the state space of
$\pi(k, \boldsymbol{\theta}_k,\boldsymbol{v}_k|\Data)=
\pi(k, \boldsymbol{\theta}_k|\Data)\tau_k(\boldsymbol{v}_k)$ is of constant
dimension for all models ${\cal M}_k\in{\mathcal M}$. 
 By updating $\boldsymbol{v}_k$ via a (deliberately) slowly
mixing Markov chain,
a
temporal memory is induced that persists in the $\boldsymbol{v}_k$ from state
to state. In this manner, the motivation behind the auxiliary variables is
to improve between-model proposals, in that some
memory of previous model states is retained.  \citet{brooks+gr03}
demonstrate that this approach can significantly enhance mixing
compared to an unassisted reversible jump sampler.
Although the fixed
dimensionality of $(k,\boldsymbol{\theta}_k,\boldsymbol{v}_k)$ is later relaxed, there is an obvious analogue 
with product space sampling frameworks \citep{carlin+c95,godsill01} -- see Section \ref{section:productspace}.

An alternative augmented state space modification of standard MCMC is given by \citet{liu+lw01}.
The dynamic weighting algorithm 
augments the original state space by a weighting factor, which permits the Markov chain to make large transitions not allowable by the standard transition rules, subject to the computation of the correct weighting factor.
Inference is then made
by using the weights to compute importance sampling estimates rather than simple Monte Carlo estimates.
This method can be used within the reversible jump algorithm to facilitate cross-model jumps.


\subsection{Local proposals in ordinal and unordered model spaces}
Some approaches that are shown to improve efficiency in MCMC over discrete spaces are applicable to sampling over multiple models. \citet{diaconis+hn00} and \citet{chen_lifting_1999} formulate a ``nearly-reversible'' method (also called ``lifting'') which introduces persistent movement in a discrete random variable with demonstrated improvements in mixing. \citet{gagnon_nonreversible_2021} apply this approach to RJMCMC proposals in nested models, i.e.~those where the model indicator $k$ is an ordinal discrete random variable, such as in change point or clustering models. The approach augments the state space with a deterministic direction variable $v\in\{-1,1\}$ such that the model space exploration is determined by $k'\gets k+v$ instead of being randomly chosen. The direction variable then alternates via $v_t\gets -v_{t-1}$ whenever a model switch is proposed.

When there is no clear ordering of models $\mathcal{M}_k$, another approach dubbed \textit{locally-balanced} proposals, initially introduced for local MCMC proposals on discrete spaces by \citet{zanella2020informed}, is applicable to RJMCMC proposals by treating the target marginal model distribution $\pi(k |\Data)$ as the discrete space on which local proposals are designed. The proposal design is 
\begin{equation}\label{eqn:informedrjmcmcproposal}
q(k'\rightarrow k) = h\left( \frac{\widehat{\pi}(k'|\Data)}{\widehat{\pi}(k|\Data)} \right),
\end{equation}
where $h$ is a user-specified function. By choosing $h$ to be the identity, the proposal reduces to the standard \textit{globally-balanced} approach, but by choosing $h=x/(1+x)$ (what the authors call the Barker proposal) or $h=\sqrt{x}$, the authors showed that the resulting Markov chain has better mixing properties. This approach requires either knowledge of or an approximation to $\pi(k'|\Data)/\pi(k|\Data)$, which can be obtained via Laplace's method.



\subsection{Multi-step proposals}

\citet{green+m01} introduce a procedure for learning from rejected between-model proposals 
based on an extension of the splitting rejection idea of \citet{tierney+m99}. After rejecting a between-model
proposal, the procedure makes a second proposal, usually under a modified proposal mechanism, and
potentially dependent on the value of the rejected proposal. In this manner, a limited form of adaptive behaviour may be incorporated into the proposals. 
Delayed-rejection schemes can reduce the asymptotic variance of ergodic averages by reducing the probability of the chain remaining in the same state \citep{peskun73,tierney98}, however there is an obvious trade-off with the extra move construction and computation required.

For clarity of exposition, in the remainder of this Section we denote  the current state of the Markov chain in model ${\cal M}_k$ by $\boldsymbol{x}=(k, \boldsymbol{\theta}_k)$, and the
first and second stage proposed states in model ${\cal M}_{k'}$ by $\boldsymbol{y}$ and $\boldsymbol{z}$. Let $\boldsymbol{y}=g^{(1)}_{k \rightarrow k'}(\boldsymbol{x}, \boldsymbol{u_1})$ and $\boldsymbol{z}=g^{(2)}_{k \rightarrow k'}(\boldsymbol{x}, \boldsymbol{u_1}, \boldsymbol{u_2})$ be the mappings
of the current state and random vectors $\boldsymbol{u_1}\sim q_{d_{k\rightarrow k'}}^{(1)}(\boldsymbol{u}_1)$ and $\boldsymbol{u_2}\sim q_{d_{k\rightarrow k'}}^{(2)}(\boldsymbol{u}_2)$ into the proposed new states. For simplicity, we again consider the framework where the dimension of model
${\cal M}_k$ is smaller than that of model ${\cal M}_{k'}$ (i.e~$n_{k'}>n_k$)  and where the reverse move proposals are deterministic. 
The proposal from $\boldsymbol{x}$ to $\boldsymbol{y}$ is accepted with the usual acceptance probability 
$$
\alpha_1(\boldsymbol{x}, \boldsymbol{y}) = \min \left\{1, \frac{\pi(\boldsymbol{y})q(k'\rightarrow k)}{\pi(\boldsymbol{x})q(k \rightarrow k')q_{d_{k\rightarrow k'}}^{(1)}(\boldsymbol{u_1})} \left | \frac{\partial g^{(1)}_{k \rightarrow k'}(\boldsymbol{x}, \boldsymbol{u_1})}{\partial (\boldsymbol{x}, \boldsymbol{u_1})}\right |\right\}.
$$
If $\boldsymbol{y}$ is rejected, detailed balance for the move from $\boldsymbol{x}$ to $\boldsymbol{z}$ is preserved with the acceptance probability 
$$
\alpha_2(\boldsymbol{x}, \boldsymbol{z}) = \min \left\{1, \frac{\pi(\boldsymbol{z})q(k'\rightarrow k)[1-\alpha_1(\boldsymbol{y}^*, \boldsymbol{z})^{-1}]}{\pi(\boldsymbol{x})q(k \rightarrow k')q_{d_{k\rightarrow k'}}^{(1)}(\boldsymbol{u_1})q_{d_{k\rightarrow k'}}^{(2)}(\boldsymbol{u_2})[1-\alpha_1(\boldsymbol{x}, \boldsymbol{y})]} \left | \frac{\partial g^{(2)}_{k \rightarrow k'}(\boldsymbol{x}, \boldsymbol{u_1}, \boldsymbol{u_2})}{\partial (\boldsymbol{x}, \boldsymbol{u_1}, \boldsymbol{u_2})}\right |\right\},
$$
where $\boldsymbol{y}^* = g^{(1)}_{k \rightarrow k'}(\boldsymbol{z}, \boldsymbol{u_1})$.
Note that the second stage proposal
$\boldsymbol{z}=g^{(2)}_{k \rightarrow k'}(\boldsymbol{x}, \boldsymbol{u_1}, \boldsymbol{u_2})$ is permitted to depend on the rejected first stage proposal $\boldsymbol{y}$ (a function of $\boldsymbol{x}$ and $\boldsymbol{u_1}$).

In a similar vein, \citet{al-awadhi+hc02} also acknowledge that an initial between-model proposal $\boldsymbol{x}'=g_{k \rightarrow k'}(\boldsymbol{x}, \boldsymbol{u})$ may be poor, and seek to adjust the state $\boldsymbol{x}'$ to a region of higher posterior probability before taking the decision to accept or reject the proposal.
Specifically,  \citet{al-awadhi+hc02} propose to initially evaluate the proposed move to $\boldsymbol{x}'$ in model ${\mathcal M}_{k'}$ through a density $\pi^*(\boldsymbol{x}')$ rather than the usual $\pi(\boldsymbol{x}')$. 
The authors suggest taking $\pi^*$ to be some tempered distribution $\pi^* = \pi^{\gamma}$, $\gamma>1$, such that the modes of $\pi^*$ and $\pi$ are aligned. 

The algorithm then implements $\kappa\geq 1$ fixed-dimension MCMC updates, 
generating states $\boldsymbol{x}'\rightarrow\boldsymbol{x}^1\rightarrow\ldots\rightarrow\boldsymbol{x}^{\kappa} = \boldsymbol{x}^*$, with each step
satisfying detailed balance with respect to $\pi^*$. This provides an opportunity for $\boldsymbol{x}^*$ to move closer to the mode of $\pi^*$ (and therefore $\pi$) than $\boldsymbol{x}'$.
The move from $\boldsymbol{x}$ in model ${\mathcal M}_k$ to the
final state $\boldsymbol{x}^*$ in model ${\mathcal M}_{k'}$ (with density $\pi(\boldsymbol{x}^*)$) is finally accepted with probability 
$$
\alpha(\boldsymbol{x}, \boldsymbol{x}^*)= \min \left\{1, \frac{\pi(\boldsymbol{x}^*)\pi^*(\boldsymbol{x}')q(k' \rightarrow k)}{\pi(\boldsymbol{x})
\pi^*(\boldsymbol{x}^*)q(k \rightarrow k') q_{d_{k\rightarrow k'}}(\boldsymbol{u})}
\left | \frac{\partial g_{k \rightarrow k'}(\boldsymbol{x}, \boldsymbol{u})}{\partial (\boldsymbol{x}, \boldsymbol{u})}\right |
\right\}.
$$
The implied reverse move from model ${\mathcal M}_{k'}$ to model model ${\mathcal M}_{k}$ is
conducted by taking the $\kappa$ moves with respect to $\pi^*$ first, followed by the 
dimension changing move.

Various extensions can easily be incorporated into this framework, such as using a sequence of $\pi^*$ distributions, resulting in a slightly modified acceptance probability expression. For instance, the standard simulated annealing framework, \citet{kirkpatrick84},  
provides an example of a sequence of distributions which encourage moves towards posterior mode. Clearly the choice of the distribution
$\pi^*$ can be crucial to the success of this strategy. As with all multi-step proposals,  increased computational overheads are traded for potentially enhanced between-model mixing.

\section{Convergence assessment}
 \label{sec:converge}

Under the assumption that an acceptably
efficient method of constructing a reversible jump sampler is
available, one obvious pre-requisite to inference is that the
Markov chain converges to its equilibrium state. 
Even in fixed dimension problems,
theoretical convergence bounds can be difficult to generalise \citep{herbertj2001, rosenthal1995}.  In the absence of such theoretical results, convergence diagnostics based on 
empirical statistics computed from the sample path of multiple chains are often the only available tool.
An obvious drawback of the empirical approach is that such diagnostics invariably fail 
to detect a lack of convergence when parts of the target distribution are missed entirely by all replicate chains. Accordingly, these are necessary rather than sufficient indicators of chain convergence.
See  \citet{cowles+c96}, \citet{roy2020}, \citet{flegalg2016}, \citet{vatsfj2019}  for comparative reviews and some recent advances under fixed 
dimension MCMC. 

The reversible jump sampler generates additional problems in the design of suitable 
empirical diagnostics, since most of these depend on the identification of suitable scalar statistics 
of the parameters sample paths. However, in the multi-model case, 
these statistics may no longer retain the same 
interpretation. In addition, 
convergence is not only required within each
of a potentially large number of models, but also across models
with respect to posterior model probabilities.

One obvious approach would be the implementation of independent
sub-chain assessments, both within-models and for the model indicator $k\in{\mathcal K}$.
With focus purely on model selection, \citet{brooks+gp03} propose various
diagnostics based on the sample-path of the model indicator, $k$,
including non-parametric hypothesis tests such as the $\chi^2$ and
Kolmogorov-Smirnov tests. In this manner,
distributional assumptions of the models (but not the statistics)
are circumvented at the price of associating marginal convergence of $k$ with convergence of the full
posterior density.

\citet{brooks+g00} propose
the monitoring of functionals of parameters which retain their interpretations 
as the sampler moves between models.  The deviance is suggested as a default choice in the absence of superior alternatives.
A two-way ANOVA decomposition of the
variance of such a functional is formed over multiple chain
replications, from which the potential scale reduction factor (PSRF) \citep{gelman+r92} can be constructed and monitored. 
\citet{castelloe+z02} extend this approach firstly to an unbalanced (weighted) two-way ANOVA, to prevent the PRSF being dominated by a few visits to rare models, with the weights being specified in proportion to the frequency of model visits.
\citet{castelloe+z02} also extend their diagnostic to the
multivariate (MANOVA) setting on the observation that monitoring
several functionals of marginal parameter subsets is more robust
than monitoring a single statistic. 
This general method is clearly reliant on the identification of useful statistics to monitor, but is also sensitive to the 
extent of
approximation induced by violations of the ANOVA assumptions of
independence and normality.

\citet{sisson+f04a} propose diagnostics when the underlying model can be formulated in the marked point process framework \citep{stephens00,diggle83}.  
For example, a mixture of an unknown number of univariate normal densities (Equation \ref{eqn:mixture}) can be represented as a set of $k$ events $\xi_{j}=(w_{j},
\mu_{j},\sigma^2_{j})$, $j=1,\ldots,k$, in a region $A\subset{\mathcal R}^3$. Given a reference point $v\in A$,
in the same space as the events $\xi_j$ (e.g.~$v=(\omega, \mu,\sigma^2)$), 
%
%
then the point-to-nearest-event distance, 
$y$, is the 
distance from the point ($v$) to the nearest event ($\xi_j$) 
in $A$ with respect to 
some distance measure. One can evaluate distributional aspects of the events $\{\xi_{j}\}$, through $y$, as
observed from different reference points $v$. A diagnostic can then be constructed based on
comparisons between empirical distribution functions of the distances $y$, constructed from Markov chain sample-paths. 
Intuitively, as the Markov chains converge, the distribution functions for $y$ constructed from
replicate chains should be similar.


This approach permits the direct comparison of full parameter
vectors of varying dimension and, as a result, naturally
incorporates a measure of across model convergence. Due to the
manner of their construction, \citet{sisson+f04a} are able to
monitor an arbitrarily large number of such diagnostics. However, while
this approach may have some appeal, it is limited by the need to
construct the model in the marked point process setting. Common models which may be formulated in this framework include
finite mixture, change point and regression models.

\noindent{\bf Example: Convergence assessment for finite mixture univariate Normals}\\
We consider the reversible jump sampler of   \citet{richardson+g97} implementing a finite mixture of Normals model (Equation \ref{eqn:mixture}) using the
enzymatic activity dataset (Figure \ref{fig:enzyme}(b)). For the purpose of assessing performance of the sampler, we implement five independent sampler replications of length 400,000 iterations.

\begin{figure}[htb]
\psfrag{Iteration (thousands)}[b][b]{{\tiny Iterations (thousands)}} 
\psfrag{P-Value}[l][l]{{\tiny P-Value}} 
\psfrag{KS statistic}[t][t]{{\tiny KS}}  
\psfrag{chisq statistic}[t][t]{{\tiny $\chi^2$}}  
\psfrag{Vhat/Wc and Wm/WmWc (Deviance)}[t][t]{{\tiny mPSRF}}  
\psfrag{Gelman and Rubin}[t][t]{{\tiny PSRFv}}
\begin{center}
\begin{subfigure}[]{0.45\textwidth}
\rotatebox{-90}{
\includegraphics[height=6.75cm,width=5.5cm]{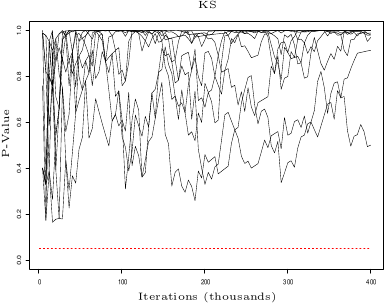} }
\caption{}
\end{subfigure} 
\begin{subfigure}[]{0.45\textwidth}
\rotatebox{-90}{
\includegraphics[height=6.75cm,width=5.5cm]{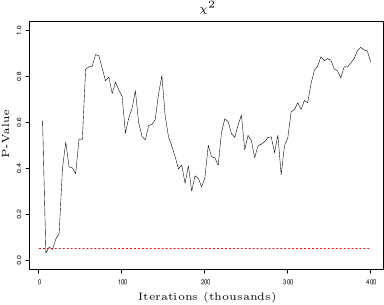}} 
\caption{}
\end{subfigure} 
\begin{subfigure}[]{0.45\textwidth}
\rotatebox{-90}{
\includegraphics[height=6.75cm,width=5.5cm]{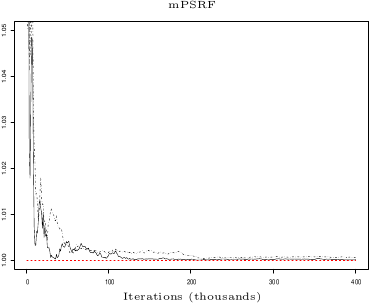}}
\caption{}
\end{subfigure}
\begin{subfigure}[]{0.45\textwidth}
\rotatebox{-90}{
\includegraphics[height=6.75cm,width=5.5cm]{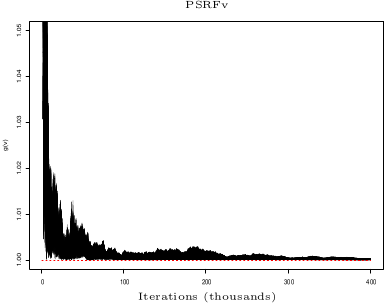}}
\caption{}
\end{subfigure}
\caption{\small Convergence assessment for the enzymatic activity dataset. Plots (a) 
Kolmogorov-Smirnov and (b) $\chi^{2}$ tests of  \citet{brooks+gp03}.  Horizontal line denotes an $\alpha=0.05$ significance level for test of different sampling distributions. Plots (c) multivariate PSRF's of \citet{castelloe+z02} and (d) PSRFv's of \citet{sisson+f04a}. Horizontal lines denote the value of each statistic under equal sampling distributions.}\label{image:otherstats-enzyme}
\end{center}
\end{figure}

Figure \ref{image:otherstats-enzyme} (a,b) illustrates the diagnostic of \citet{brooks+gp03} which provides a test for between-chain convergence based on posterior model probabilities. 
The pairwise Kolmogorov-Smirnov and $\chi^{2}$ (all chains simultaneously)
tests assume independent realisations.
Based on the estimated convergence rate, \citet{brooks+gp03},
we retain every 400th iteration to obtain approximate independence. 
The Kolmogorov-Smirnov statistic cannot reject immediate convergence, with all pairwise chain
comparisons well above the critical value of 0.05. The 
$\chi^{2}$ statistic cannot reject convergence after the first 10,000 iterations.

Figure \ref{image:otherstats-enzyme} (c) illustrates the two multivariate PSRF's of \citet{castelloe+z02} using the deviance as the default statistic to monitor. 
The solid line shows the ratio of between- and within-chain variation;
the broken line indicates the ratio of 
within-model variation, and the within-model, within-chain variation. 
The mPSRF's 
rapidly approach 1, suggesting convergence,
beyond 166,000 iterations. This is supported by the independent analysis 
of \citet{brooks+g00} who demonstrate evidence for 
convergence of this sampler after around 150,000 iterations,
although they caution that their chain lengths of only 200,000 iterations were too 
short for certainty.

Figure \ref{image:otherstats-enzyme} (d), adapted from \citet{sisson+f04a}, illustrates the PSRF of the distances from each of 100 randomly chosen reference points to the nearest model components, over the five replicate chains. 
Up to around 100,000 iterations,  between-chain variation is still reducing; beyond 300,000 iterations, 
differences between the chains appear to have stabilised. The intervening iterations mark a gradual transition between these two states. This diagnostic appears to be the most conservative of those presented here.


This example highlights 
that 
empirical convergence assessment tools 
often give varying estimates of when convergence may have been achieved.
As a result, it may
be prudent to follow the most conservative estimates in practice. 
While it is undeniable that the benefits for the practitioner in
implementing reversible jump sampling schemes are immense, it
is arguable that the practical importance of ensuring
chain convergence is often overlooked. 
However, it is also likely that current diagnostic methods are insufficiently
advanced to permit a more rigorous default assessment of sampler
convergence.

\section{Model choice and Bayes factors}\label{sec:BF}

Bayesian model selection is canonically implemented using estimates of Bayes Factors \citep{kass+r95}. It is usually the case that more than one model provides useful statistical inference, and in such cases one can take expectations against a collection of models, weighted by their posterior probabilities. This is known as \emph{Bayesian model averaging} \citep{hoeting+mrv99} where, given a quantity of interest $\Delta$, the posterior given data $\Data$ is
\[
\pi(\Delta|\Data)=\sum_{k\in\mc{K}}\pi(\Delta|k,\Data)\pi(k|\Data)\label{eq:bgbmaposterior},
\]
which is the average of the conditional posteriors of $\Delta$ weighted by the posterior model probabilities $\pi(k|\Data)$.

One of the useful by-products of the reversible jump sampler, is the ease with which Bayes factors
can be estimated. Explicitly expressing marginal or predictive densities of $\Data$ 
under model ${\cal M}_k$ as
$$
m_k(\Data)=\int_{{\cal R}^{n_k}}
L(\Data|k, \boldsymbol{\theta}_k)p(\boldsymbol{\theta}_k| k)d\boldsymbol{\theta}_k,
$$
the normalised posterior probability of model ${\cal M}_k$ is
given by
\begin{eqnarray*}
\label{eqn:posterior_prob} 
p(k|\Data)=
\frac{p(k)m_k(\Data)}{ \sum_{k'\in{\mathcal K}}p(k')m_{k'}(\Data)} = \left(1+\sum_{k'\neq
k}\frac{p(k')}{p(k)}B_{k',k}\right)^{-1},
\end{eqnarray*}
where $B_{k',k}=m_{k'}(\Data)/m_k(\Data)$ is the
Bayes factor of model ${\cal M}_{k'}$ to ${\cal M}_k$, and $p(k)$ is the prior
probability of model ${\mathcal M}_k$.  For a
discussion of Bayesian model selection techniques, see
\citet{chipman+gm01}, \citet{berger+p01}, \citet{kass+r95},
\citet{ghosh+t01}, \citet{berger+p04}, \citet{barbieri+b04}. 
A usual estimator of the posterior model probability, $p(k|\Data)$, is given by the proportion of chain iterations the reversible jump sampler spent in model ${\cal M}_k$.

\subsection{Bayes factor via reversible jump}
When the number of candidate models $|{\cal M}|$ is large,
the use of reversible jump MCMC algorithms to evaluate Bayes
factors raises issues of efficiency. 
Suppose that model ${\cal M}_k$ accounts for a large proportion of posterior mass.
In attempting a between-model move from
model ${\cal M}_k$,
the
reversible jump algorithm will tend to persist in this model 
and visit others models rarely. Consequently, 
estimates of Bayes factors based on model-visit proportions will
tend to be inefficient \citep{han+c01}. 

\citet{bartolucci+sm06} propose enlarging the parameter space of the models under comparison
with the same auxiliary variables, $\boldsymbol{u}\sim q_{d_{k\rightarrow k'}}(\boldsymbol{u})$ and $\boldsymbol{u}' \sim q_{d_{k'\rightarrow k}}(\boldsymbol{u}')$ (see Equation \ref{eqn:dmatching2}), defined under the between-model transitions, so that the enlarged spaces, $(\boldsymbol{\theta}_k,\boldsymbol{u})$ and $(\boldsymbol{\theta}_{k'},\boldsymbol{u}')$, have the same dimension.
In this setting, an extension to the Bridge estimator for the estimation of the ratio of normalising constants
of two distributions \citep{meng+w96} can be used, 
by integrating out the auxiliary random process (i.e. $\boldsymbol{u}$ and $\boldsymbol{u}'$) involved in the between-model moves.
Accordingly, the Bayes factor of model ${\mathcal M}_{k'}$ to ${\mathcal M}_k$ can be estimated using the reversible jump acceptance probabilities as
$$
\hat{B}_{k',k} = \frac{\sum_{j=1}^{J_k}\alpha^{(j)}[(k, \boldsymbol{\theta}_k),(k',  \boldsymbol{\theta}'_{k'})]/J_k}{\sum_{j=1}^{J_{k'}}\alpha^{(j)}[(k', \boldsymbol{\theta}'_{k'}),(k,  \boldsymbol{\theta}_k)]/J_{k'}}
$$
where $\alpha^{(j)}[(k, \boldsymbol{\theta}_k),(k',  \boldsymbol{\theta}'_{k'})]$ is the acceptance probability (Equation \ref{eqn:dmatching2}) of the $j$-th attempt to move from model ${\mathcal M}_k$ to ${\mathcal M}_{k'}$, and where $J_k$ and $J_{k'}$ are the number of proposed moves from model ${\mathcal M}_k$ to ${\mathcal M}_{k'}$ and vice versa  during the
simulation. Further manipulation is required to estimate $B_{k',k}$ if the sampler does not jump between models ${\mathcal M}_k$ and ${\mathcal M}_{k'}$ directly \citep{bartolucci+sm06}. This approach can provide a more efficient way of postprocessing reversible jump MCMC with minimal computational effort.

\subsection{Bayes factors via transdimensional annealed importance sampling}

An alternative approach developed by \citet{karagiannis_annealed_2013} adopts the annealed importance sampling paradigm to generate a path between $\mathcal{M}_k$ and $\mathcal{M}_{k'}$ that will yield the Bayes factor estimate $\widehat{B}_{k',k}$ . It is a natural extension to apply a resampling step in the vein of sequential Monte Carlo (SMC), as discussed in \citet{zhou_toward_2016} and explored further in \citet{everitt_sequential_2020}.

For ease of exposition, we adopt notation to decompose the diffeomorphism $g_{k\mapsto k'}$ into constituent components $\bm{\theta}_{k'}' \gets g_{k\mapsto k'}^{\bm{\theta}}(\bm{\theta}_k,\bm{u})$ and $\bm{u}' \gets g_{k\mapsto k'}^{\bm{u}}(\bm{\theta}_k,\bm{u})$. For a given sequence of monotonically increasing temperatures $0=\gamma_0<\dots<\gamma_T=1$, the unnormalised annealed target distribution is
\begin{multline*}
\eta_{t,k\mapsto k'}(\bm{\theta}_{k'}',\bm{u}') = \bigg[\pi(\bm{\theta}_{k'}|k', \Data),q_{d_{k'\mapsto k}}(\bm{u}')\bigg]^{\gamma_t}\\
\left[\pi\big(g_{k'\mapsto k}^{\bm{\theta}}(\bm{\theta}_{k'} |k, \Data\big)q_{d_{k\mapsto k'}}\big(g_{k'\mapsto k}^{\bm{u}}(\bm{u}')\big)\left|\dfrac{\partial g_{k'\mapsto k}(\bm{\theta}_{k'}',\bm{u}')}{(\bm{\theta}_{k'}',\bm{u}')}\right|\right]^{1-\gamma_t}.
\end{multline*}
Given a proposed model $k'\sim q$, particles are transformed via $\bm{\theta}_{k'}', \bm{u}'\gets g_{k\mapsto k'}(\bm{\theta}_k,\bm{u})$, $\bm{u}\sim q_{d_{k\mapsto k'}}$. For notational convenience, we write the incremental Bayes factor estimate at step $t$ as $\widehat{B}_{t,k\mapsto k'}$. For the initial temperature $\gamma_0=0$, initial normalised weights of particles are set to $W_0^{(i)}=N^{-1}$ $\forall i=1,\ldots, N$ and the initial 
 Bayes factor estimate is set to $\widehat{B}_{0,k\mapsto k'}=1$. Then, over the sequence of temperatures $\{\gamma_t\}_{t=1}^{T}$ the weight update for the $i^{\mathrm{th}}$-particle is
\[
w_{t}^{(i)} = W_{t-1}^{(i)}\dfrac{\eta_{t,k\mapsto k'}(\bm{\theta}_{k'}'^{(i)},\bm{u}'^{(i)})}{\eta_{t-1,k\mapsto k'}(\bm{\theta}_{k'}'^{(i)},\bm{u}'^{(i)})}.
\]
After updating each weight for step $t$, the Bayes factor estimate is updated to be
\[
\widehat{B}_{t,k\mapsto k'} \gets \widehat{B}_{t-1,k\mapsto k'}\sum_{i=1}^N w_{t}^{(i)}.
\]
Weights are then normalised via $W_t^{(i)}\gets w_t^{(i)}/\sum_{j=1}^N w_t^{(j)}$, and for the SMC variants of this annealing procedure, resampling is conducted using these normalised weights. Lastly, particles are diversified via a $\eta_{t,k\mapsto k'}$-invariant MCMC kernel before incrementing $t\gets t+1$ and repeating for the remaining $\gamma_{\geq t}$ temperatures.



\section{Multi-model sampling: beyond reversible jump}
\label{sec:related}


Several alternative multi-model sampling methods are available. Some of these are closely related to the reversible jump MCMC algorithm, or include reversible jump as a special case.

\subsection{Transdimensional piecewise deterministic Markov processes}


MCMC methods canonically operate by obtaining point samples of a target distribution. An alternative to this approach, called piecewise deterministic Markov processes (PDMPs) \citep{davis1984piecewise,costa2008stability}, instead characterises a target distribution $\pi$ on support $\mc{Z}=\mc{X}\times\mc{V}$, where $\mc{X}\subseteq\mathcal{R}^d$ and $\mc{V}$ is a space of auxiliary variables, using deterministic \emph{trajectories} (or flows), denoted $\phi(dt,\bm{z})$ where $\bm{z}\in\mc{Z}$ is the initial state and $t$ is time. A piecewise deterministic Markov process $Z(t)$ is defined by the choice of $\phi$, a set of random times $T_1,T_2,\dots$ at which the process jumps (usually exponentially distributed with rate $\lambda(Z(t))$), and finally a measure $Q(\bm{z},d\bm{z'})$ which defines how the process moves from $\bm{z}\in\mc{Z}$ to $\bm{z'}\in\mc{Z}$ at each jump time. The key feature is that the dynamics of $\phi$ are deterministic between jumps, whereby simulation from the PDMP is generally
\begin{align*}
    Z(t) &= \phi(t - T_{i}, Z(T_{i})), \quad \text{for} \quad T_{i} \leq t < T_{i+1} \\
    Z(T_{i+1}) &\sim Q(Z(T_{i}), \,\cdot\,).
\end{align*}
A popular PDMP sampler is the Zig-Zag process \citep{10.1214/18-AOS1715}, denoted $Z(t)=(X(t),\varphi(t))$ and defined on the augmented state space $\mc{Z}=\mc{X}\times\{-1,1\}^d$, $\mc{X}\subseteq\mc{R}^d$, that incorporates a ``velocity'' $\varphi(t)\in\{-1,1\}^d$. The jump mechanism $Q((\bm{x}, \varphi), (d\bm{x'}, d\varphi')) = \delta_{\bm{x}}(d\bm{x'}) \times \delta_{\text{Flip}(\varphi)}(d\varphi')$ component-wise flips the sign of the velocity at the jump time. The jump rate is $\lambda(\bm{x}, \varphi(t)) = \mathrm{max}(0,-\varphi(t) \cdot \nabla \log \pi(\bm{x}))$, effectively ensuring that the process reflects off the level sets of $\pi$. 

Motivated by the application of PDMPs to transdimensional problems such as variable selection, where the support of $\pi$ over all models $\mc{M}_1,\mc{M}_2,\dots,$ indexed by $k$, is $\mc{X}=\cup_{k}\mc{X}_k$, \citet{chevallier_reversible_2022} present a reversible jump formulation that naturally extends the piecewise deterministic approach with reversible deterministic transitions between models. By way of example, we will examine a reversible jump Zig-Zag (RJZZ) process on a variable selection model, where a jump between models is written such that model $\mc{M}_j$ is obtained by removing one variable from the support of model $\mc{M}_i$. In this case, the RJZZ process has a between-model jump mechanism $Q_{j,i}$ that is triggered when a trajectory $Z_i(t)$ (the process in model $\mc{M}_i$) intersects with a zero axis. The process will jump to the model $\mc{M}_j$ (the model with this variable removed) via $Q_{j,i}$ setting the velocity to zero for this variable, which causes it to remain at zero and for the process to stay in model $\mc{M}_j$. The velocity for this variable is reintroduced by simulating uniformly from $\{-1,1\}$, and the rate at which a component velocity is reintroduced follows similar conditions to the RJMCMC framework.

\begin{figure}
    \centering
    \includegraphics[width=\textwidth]{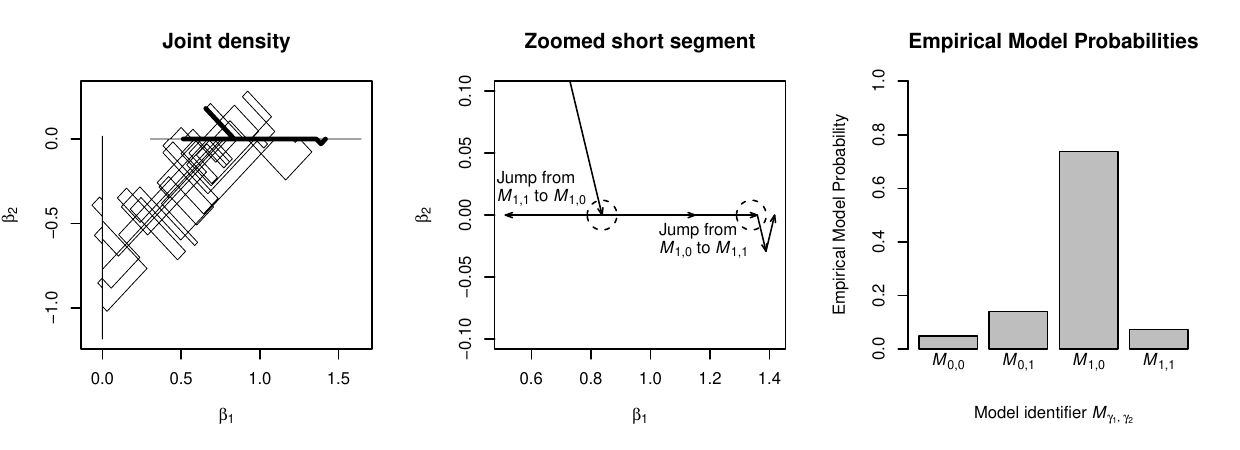}
    \caption{\small \textit{Left}: Joint posterior samples for a reversible-jump Zig-Zag process \citep{chevallier_reversible_2022} run for 500 iterations on a 2-variable logistic model, where $\beta_1,\beta_2$ denotes the coefficient of the two variables. When the process is in model $\mc{M}_{1,0}$ or model $\mc{M}_{0,1}$, the process samples  1D as visualised by the horizontal and vertical lines respectively, and model $\mc{M}_{1,1}$ is visualised by the 2D zig-zag trajectory. \textit{Centre}: A zoomed segment of 6 piecewise trajectories from the joint density (shown in \textbf{bold} in the left plot) showing the jumps from models $\mc{M}_{1,1}$ to $\mc{M}_{1,0}$. \textit{Right}: The empirical model probabilities of each model ($\mc{M}_{0,0}$, $\mc{M}_{0,1}$, $\mc{M}_{1,0}$, and $\mc{M}_{1,1}$) derived from the time (i.e.~length of the trajectory) the process spends in each model.}
    \label{fig:rjpdmpzigzag}
\end{figure}

Since the piecewise trajectories are continuous-time, the RJZZ process will hit zero exactly with a probability of 1 for variables with low support, making reversible moves over models a much more straightforward process than for Hamiltonian Monte Carlo and other gradient-based samplers where the discrete leapfrog trajectory will skip over the zero axis. 
Figure \ref{fig:rjpdmpzigzag} shows an example of the RJZZ process on a 2-variable logistic regression model, where the competing models are denoted by $\mathcal{M}_{\gamma_1,\gamma_2}$, where $\gamma_1,\gamma_2\in (0,1)$ are variable inclusion  indicators such that for example $\mathcal{M}_{0,1}$ denotes the model with only the second covariate, and so forth. 

%

A related method, called the sticky PDMP \citep{bierkens_sticky_2023}, differs from the reversible jump PDMP approach by allowing non-reversible model jumps. For the above variable selection scenario, the main difference is that the sticky PDMP sampler remembers the velocity of a component when it is re-introduced back into the current state rather than randomly sampling it.


\subsection{Jump diffusion}
Before the development of the reversible jump sampler,
\citet{grenander+m94} proposed a sampling strategy based on
continuous time jump-diffusion dynamics.
This process  combines jumps between models at random
times, and within-model updates based on a diffusion process according to
a Langevin stochastic differential equation indexed by time, $t$, satisfying
\begin{equation*}
\label{eqn:diffusion}
d\boldsymbol{\theta}_k^t = dB_k^t + \frac{1}{2}\nabla\log\pi(\boldsymbol{\theta}_k^t, k|\Data)dt
\end{equation*}
where $dB_k^t$ denotes an increment of Brownian motion, and $\nabla$
the vector of partial derivatives.  

The probability of jumping out of a model $\mathcal{M}_k$, is specified through a jump intensity $q((\boldsymbol{\theta}_k,k) \rightarrow (\boldsymbol{\theta}'_{k'}, k'))$. To decide when to jump, the marginal jump intensity is calculated (marginalising over $\boldsymbol{\theta}'_{k'}$ and $k'$) and the random jump times can be sampled by generating unit exponential random variables. Detailed balance conditions are satisfied by choosing the appropriate jump intensity to ensure the correct target for the stationary distribution.

This method has found some application in signal processing and other Bayesian
analyses \citep{miller+sg95,phillips+s96}, but has in general been superceded by the more
accessible reversible jump sampler. 
In practice, the continuous-time diffusion must be approximated by a discrete-time
simulation. If the  time-discretisation is corrected for via a
Metropolis-Hastings acceptance probability, the
jump-diffusion sampler 
actually results in an implementation of reversible jump MCMC \citep{besag94}. 

Recently in machine learning, generative models based on diffusion processes have shown great performance on wide range of problems \citep{yang2023}. These models define a forward diffusion process that corrupts data to noise and then a backward generative process that generates new data from noise.
When the dimension of the data vary, \citet{campbell2023transdimensional} proposed a transdimensional generative model based on jump diffusions. In the forward process, a jump step destroys dimensions and in the backward step, dimensions are added by the jumps. 
\subsection{Product space formulations}
\label{section:productspace}

As an alternative to samplers designed for implementation on
unions of model spaces, $\boldsymbol{\Theta}=\bigcup_{k \in {\cal K}} (\{k\}, {\cal R}^{n_k})$,   
``super-model'' product-space frameworks have been developed,
with a state space given by $\boldsymbol{\Theta}^*=\otimes_{k\in\mathcal{K}} (\{k\},{\cal R}^{n_k})$.
This setting encompasses all model spaces jointly, so that a sampler needs to simultaneously track 
$\boldsymbol{\theta}_k$ for all $k\in{\mathcal K}$.
The composite parameter vector, $\boldsymbol{\theta}^*\in\boldsymbol{\Theta}^*$, consisting of a
concatenation of all parameters
under all models, is of fixed-dimension, thereby circumventing the necessity of between-model transitions.
Clearly, product-space samplers are limited to situations where the dimension of $\boldsymbol{\theta}^*$ is computationally feasible.
\citet{carlin+c95}
propose a
 posterior distribution for the
composite model parameter and model indicator
given by
\begin{equation*}
\label{eqn:carlin_chibb} \pi(k,\boldsymbol{\theta}^*|\Data)\propto
L(\Data| k, \boldsymbol{\theta}^*_{{\mathcal I}_k})
p(\boldsymbol{\theta}^*_{{\mathcal I}_k}| k)p(\boldsymbol{\theta}^*_{{\mathcal
I}_{-k}}|\boldsymbol{\theta}^*_{{\mathcal I}_k}, k)p(k),
\end{equation*}
where ${\mathcal I}_k$ and ${\mathcal I}_{-k}$ are index sets
respectively identifying and excluding the parameters $\boldsymbol{\theta}_k$
from $\boldsymbol{\theta}^*$. Here ${\mathcal I}_k\cap{\mathcal
I}_{k'}=\emptyset$ for all $k\neq k'$, so that the parameters for
each model are distinct. It is easy to see that the term $p(\boldsymbol{\theta}^*_{{\mathcal
I}_{-k}}|\boldsymbol{\theta}^*_{{\mathcal I}_k}, k)$, called a ``pseudo-prior" by \citet{carlin+c95},  has no effect on the joint posterior $\pi(k,\boldsymbol{\theta}^*_{{\mathcal I}_k} | \Data)=\pi(k,\boldsymbol{\theta}_k|\Data)$, and its form is usually chosen for
convenience.  However, poor choices may affect the efficiency of the sampler \citep{green03,godsill03}.

\citet{godsill01} proposes a further
generalisation of the above 
by relaxing the restriction that ${\mathcal
 I}_k\cap{\mathcal I}_{k'}=\emptyset$ for
all $k\neq k'$.
That is, individual model parameter vectors are permitted to overlap
arbitrarily,
which is intuitive for, say, 
nested models.  
This framework can be shown to encompass the reversible jump algorithm, in addition to the setting of \citet{carlin+c95}. In theory this allows for direct comparison between the three samplers, although this has not yet been fully examined.
However, one clear point
is that the information contained within $\boldsymbol{\theta}^*_{{\mathcal I}_{-k}}$
would be useful in generating efficient between-model transitions
when in model ${\cal M}_k$, under a reversible jump sampler. This idea is exploited by
\citet{brooks+gr03}.

\subsection{Point process formulations}

A different perspective on the multi-model sampler is based
on spatial birth-and-death processes
\citep{preston77,ripley77}.
\citet{stephens00} observed that
particular 
multi-model statistical problems
can be represented
as continuous time, marked point processes \citep{geyer+m94}. 
\citep[The RJMCMC convergence diagnostic of][is directly applicable here; see Section \ref{sec:converge}]{sisson+f04a}.
One obvious setting is finite mixture modelling (Equation \ref{eqn:mixture}) where 
the birth and death of mixture components, $\boldsymbol{\phi}_j$, indicate transitions between models.  
The sampler of
\citet{stephens00} may be interpreted as a particular continuous time,
limiting version of a sequence of reversible jump algorithms  \citep{cappe+rr03}.

A number of illustrative comparisons of the reversible jump, jump-diffusion, product space and point process frameworks can
be found in the literature. 
See, for example, \citet{andrieu+dd01},
\citet{dellaportas+fn02}, \citet{carlin+c95}, \citet{godsill01,godsill03}, \citet{cappe+rr03} and \citet{stephens00}.

\subsection{Multi-model optimisation}

The reversible jump MCMC sampler may be utilised as the underlying random mechanism within a stochastic optimisation framework, 
given its ability to traverse complex spaces efficiently \citep{brooks+fk03,andrieu+fd00}.
In a simulated annealing setting, the sampler would define a stationary
distribution proportional to the
Boltzmann distribution
\begin{eqnarray*}
\mathcal{B}_T(k, \boldsymbol{\theta}_k)\propto
\exp\{-f(k, \boldsymbol{\theta}_k)/T\},
\end{eqnarray*}
where $T\geq 0$  and $f(k, \boldsymbol{\theta}_k)$,  is
a model-ranking function to be minimised. A stochastic annealing
framework will then 
decrease the value
of $T$ according to some schedule while using the reversible jump sampler to
explore function space. Assuming adequate chain mixing, as $T\rightarrow 0$ the sampler and the Boltzmann distribution
will converge to a point mass at
$(k^*, \boldsymbol{\theta}^*_{k^*})=\arg\max f(k,\boldsymbol{\theta}_k)$. 
Specifications for the model-ranking function may include
the
AIC or BIC \citep{sisson+f04b,king+b02}, the posterior model probability
\citep{clyde99}
or a non-standard loss function defined on variable-dimensional space \citep{sisson+h04} 
for the derivation of Bayes rules.

\subsection{Multi-model population MCMC}

The population Markov chain Monte Carlo method \citep{liang+w01,liu01} may be extended to the reversible jump setting \citep{jasra+sh07}. Motivated by simulated annealing \citep{geyer+t95}, $N$ parallel reversible jump samplers are implemented targetting 
 a sequence of related distributions $\{\pi_i\}, i=1,\ldots,N$, which may be tempered versions of the distribution of interest, $\pi_1=\pi(k,\boldsymbol{\theta}_k|\Data)$. The chains are allowed to interact, in that the states of any two neighbouring (in terms of the tempering parameter) chains may be exchanged, thereby improving the mixing across the population of samplers both within and between models.
\citet{jasra+sh07}  demonstrate superior convergence rates 
over a single reversible jump sampler.  For samplers that make use of tempering 
or parallel simulation techniques, 
\citet{gramacy+sk08} propose efficient methods of utilising samples from all distributions (i.e.~including those not from $\pi_1$) using importance weights, for the calculation of given estimators.

\subsection{Multi-model sequential Monte Carlo}

The idea of running multiple samplers over a sequence of related distributions may also considered under a sequential Monte Carlo  framework  \citep{moral+dj06}. A na\"ive implementation proceeds by simply using an RJMCMC kernel in the mutation step, as explored in \citet{andrieu_sequential_1999}, but this can result in a highly variable posterior depending on the combination of prior and intermediate distributions used. \citet{jasra+dsh08} propose implementing $N$ separate SMC samplers, each targeting a different subset of model-space. 
At some stage the samplers are allowed to interact and are combined into a single sampler. This approach permits more accurate exploration of models with lower posterior model probabilities than would be possible under a single sampler.
As with population MCMC methods, the benefits gained in implementing $N$ samplers must be weighed against the extra computational overheads.

\section{Some discussion and future directions}
\label{sec:discuss}


Given the degree of complexity associated with the implementation of
reversible jump MCMC, a major focus for future research is in designing simple, yet efficient samplers, with the ultimate goal of automation. Several authors have provided new insight on the reversible jump sampler which may contribute towards achieving such goals. For example, \citet{keith+kb04} present a generalised Markov sampler,  and in a similar vein \citet{neklyudov_involutive_2020} present a generalised ``involutive'' MCMC framework, both of which include the reversible jump sampler as a special case. \citet{petris+t03} demonstrate a geometric approach for sampling from nested models, formulated
by drawing from a fixed-dimension auxiliary continuous distribution on the largest model subspace, and then using transformations to recover model-specific samples. 

An alternative way of increasing sampler efficiency would be to explore the ideas
introduced in adaptive MCMC. As with standard MCMC, any adaptations must be implemented with care
-- transition kernels dependent on the entire history of the Markov chain can only be used under diminishing adaptation conditions \citep{roberts+r08,haario+st01}. 
Alternative schemes permit modification of the proposal distribution at regeneration times, when the next state of the Markov chain becomes completely independent of the past \citep{gilks+rs98,brockwell+k05}.
Under the reversible jump framework, regeneration can be naturally achieved by incorporating an additional model, from which independent samples can
be drawn. 
Under any adaptive scheme, however, consideration needs to be given to how best to make use of historical chain information. One approach could be the use of transports \citep{davies_transport_2023} which can be learned during an MCMC burn-in, forgoing the need for pilot runs that were previously required for adaptive proposals based on mixture models.
Additionally, efficiency gains through adaptations should naturally outweigh the costs of handling chain history and modification of the proposal mechanisms.


There has been recent interest in sampling over very large model spaces such as those used for architecture selection in Bayesian neural network models \citep{Berezowski2022}, and in the presence of very large data sets, the use of stochastic gradients in single model inference \citep{pmlr-v32-cheni14,welling2011bayesian} is yet to be fully explored in a multi-model setting. However, as an alternative to traditional sampling approaches, transdimensional PDMP methods naturally lend themselves to the use of stochastic gradients \citep{chevallier_reversible_2022, bierkens_sticky_2023} and are competitive in the context of very large model spaces.

Finally, two areas remain under-developed in the context of reversible jump simulation. The first of these is perfect simulation, which
provides an MCMC framework for producing samples exactly from the target distribution, circumventing convergence issues entirely \citep{propp+w96}. Some tentative steps have been made in this area \citep{brooks+fr02}.
Secondly, while the development of approximate Bayesian inference or ``likelihood-free'' MCMC has received much
recent attention \citep{sisson2018}, implementing the sampler in the multi-model setting remains a challenging problem, in terms of both computational efficiency and bias of posterior model probabilities.

\section*{Acknowledgments}

This work was supported by the Australian Research Council (including DP230102070) and the CSIRO Future Science Platform on Machine Learning and Artificial Intelligence.

\bibliographystyle{apalike} 
\bibliography{RJMCMC}


\end{document}